\documentclass{aa}
\usepackage{natbib}
\usepackage{hyperref}
\usepackage{graphicx}
\usepackage{txfonts}
\usepackage{marginnote}
\usepackage{color}
\begin{document} 

\newcommand{\dd}{deg$^{2}$}
\newcommand{\flux}{$\rm erg \, s^{-1} \, cm^{-2}$}
\newcommand{\LL}{$\lambda$}
\newcommand{\redmapper}{redMaPPer}

   \title{\textbf{The X-CLASS$-$redMaPPer galaxy cluster comparison}}

   \subtitle{I. Identification procedures}

   \author{T. Sadibekova\inst{1}
                  \and
          M. Pierrre\inst{1}
          \and
          N. Clerc\inst{2}, 
          L. Faccioli\inst{1}, 
          R. Gastaud\inst{3}, 
          J.-P. Le Fevre \inst{3}, 
          E. Rozo\inst{4}, 
          E. Rykoff\inst{4}}
   \institute{
   Service d'astrophysique, IRFU, CEA Saclay, France\\     
   \email{tatyana.sadibekova@cea.fr}
         \and
            Max Planck Institut f\"ur Extraterrestische Physik, Postfach 1312, 85741 Garching bei M\"unchen, Germany
         \and 
         Service d'\'electronique, des d\'etecteurs et d'informatique, IRFU, CEA Saclay, France
         \and
        KIPAC, SLAC National Accelerator Laboratory, Menlo Park, CA 94025      
             }

   \date{\today}

 
  \abstract
   {This paper is the first in a series undertaking a comprehensive correlation analysis between optically selected and X-ray-selected cluster catalogues. The rationale of the project is to develop a holistic picture of galaxy clusters utilising optical and X-ray-cluster-selected catalogues with well-understood selection functions.}
   {Unlike most of the X-ray/optical cluster correlations to date, the present paper focuses on the non-matching objects in either waveband. We investigate how the differences observed
between the optical and X-ray catalogues may stem from (1) a shortcoming of the
detection algorithms, (2) dispersion in the X-ray/optical scaling
relations, or (3) substantial intrinsic differences between the cluster
populations probed in the X-ray and optical bands. The aim is to inventory and elucidate these effects in order to account for selection biases in the further  determination of X-ray/optical cluster scaling relations.}
   {We correlated the X-CLASS serendipitous cluster catalogue extracted from the XMM archive with the redMaPPer optical cluster catalogue derived from the Sloan Digitized Sky Survey  (DR8). We performed  a detailed and, in large part, interactive analysis of the matching output from the correlation. The overlap between the two catalogues has been accurately determined  and possible cluster positional errors were manually recovered. The final samples comprise 270 and 355 redMaPPer and X-CLASS clusters, respectively. X-ray cluster matching rates were analysed as a function of optical richness. In the second step, the redMaPPer clusters were  correlated with the entire X-ray catalogue, containing point and uncharacterised sources (down to a few $10^{-15}$ \flux\ in the [0.5-2] keV band). A stacking analysis was performed for the remaining undetected optical clusters.}
   {We find that all rich ($\lambda \geq 80$) clusters are detected in X-rays out to z=0.6.  Below this redshift, the richness threshold for X-ray detection steadily decreases with redshift.  Likewise, all X-ray bright cluster are detected by \redmapper. After correcting for obvious pipeline shortcomings (about 10\% of the cases both in optical and X-ray),  $\sim$  50\% of the redMaPPer (down to a richness of 20) are found to coincide with an X-CLASS cluster; when considering X-ray sources of any type, this fraction increases to $\sim$  80\%; for the remaining objects, the stacking analysis finds a weak signal within 0.5 Mpc around the cluster optical centres. The fraction of clusters totally dominated by AGN-type emission appears to be a few percent.  Conversely, $\sim$  40\% of the X-CLASS clusters are identified with a redMaPPer (down to a richness of 20) - part of the non-matches being due to the X-CLASS sample extending further out than redMaPPer ($z<1.5$ vs $z<0.6$), but extending the correlation down to a richness of 5 raises the matching rate to $\sim$ 65\%. }
   {This state-of-the-art study involving two well-validated cluster catalogues has shown itself to be complex, and it points to a number of issues inherent to blind cross-matching, owing both to pipeline shortcomings and cluster peculiar properties. These can only been accounted for after a manual check. The combined X-ray and optical scaling relations will be presented in a subsequent article. }


   \maketitle   
%

\section{Introduction}

The abundance of galaxy clusters is a powerful cosmological probe (e.g. \cite{henry09}, \cite{vikhlinin09}, \cite{mantz10a}, \cite{rozo10}, \cite{pierre11}, \cite{clerc12a}).  Indeed, galaxy clusters have provided the first line of evidence for dark matter (\cite{zwicky33}) and evidence that the matter density of the universe was sub-critical ($\Omega_m < 1$, \cite{gott74}).

Historically, galaxy clusters were first identified in the optical (\cite{abell58}).  Early optical cluster catalogues were constructed utilising single-band photometric data and were therefore extremely susceptible to selection effects.  With the advent of the ROSAT All Sky Survey (RASS, \cite{voges99}), cluster detection was primarily pursued in the X-ray, because the detection of X-ray photons provided unambiguous evidence of a deep potential well and therefore of the reality of the detected galaxy clusters.  This led to generating a plethora of RASS X-ray catalogues (e.g. \cite{ebeling00}, \cite{bohringer00}, \cite{reiprich02}, and many others), which have since been complemented both by targeted \citet{pacaud07} and serendipitous (\cite{barkhouse06,lloyd-davies11,clerc12b,takey13}) cluster searches with the XMM-Newton or Chandra observatory.   At the same time, the advent of multi-band photometric data has led to dramatic improvements in optical cluster finding and an explosion of algorithms (e.g. \cite{gladders05}, \cite{koester07}, \cite{wen13}, \cite{hao10}, \cite{szabo11}, and many others).\\

To date, cluster searches in the X-ray, optical, and now in infrared wave band for the z>1 range are still conducted independently, although simultaneous multi-band approaches are being proposed (e.g. \cite{cohn09}, \cite{bellagambda11} - assuming basic relations between the cluster observables). These catalogues are subsequently correlated,  possibly with the goals of searching for extreme objects (e.g. \cite{andreon11}) but, more generally, for establishing a correspondence (i.e. a scaling relation) between  mass proxies, such as  X-ray gas temperature and optical richness (e.g.  \cite{popesso04, popesso05, rykoff08, gal09, wen12, takey13, rozo14}). These cross-correlations can involve up to a few thousand objects and are performed in a so-called blind way with little attention to the objects left out by the procedure. This occurs in a general astrophysical context, where the use of clusters of galaxies as cosmological probes has again come under scrutiny. Most of the criticisms concern our actual ability to perform cluster mass measurements suitable for cosmological studies - i.e. to an accuracy that matches today's precision cosmology requirements (e.g. \cite{vondenlinden14}, \cite{israel14}). The main arguments invoked are:  instrumental calibration issues (\cite{rozo14b, planck13}),  biases in hydrostatic mass estimates, reliability of the mass proxy used (e.g. is the gas mass fraction truly universal?), biases introduced by galaxy-colour selections, uncontrolled projection effects in the optical or infrared cluster searches.  \\
In parallel, recent analyses have insisted on the inability of cluster-based cosmology to be disconnected from determination of the cluster scaling relations and from a detailed account of the selection biases affecting the samples (\cite{pacaud07,mantz10b,allen11}). The three aspects are intricately related and must be handled in a self-consistent way. Even at a simpler level, for a fixed cosmology, the determination of the scaling relations must include modelling of the selection, unless the objects of interest lie well above the survey detection limits. Furthermore, one of the key parameters entering the analysis is the intrinsic scatter of the scaling relations. This quantity has a critical effect on the predicted number of detected clusters and how the samples are biased towards, for instance, more luminous objects with respect to the mean (given the steepness of the mass function). Scatter values are  hardly known in the local universe because they require large samples to be determined and, consequently, should be left as supplementary free parameters in the cosmological analyses.\\
In this context, we have undertaken an extensive correlation study between an X-ray catalogue and an optical one, namely X-CLASS extracted from the XMM all-sky archives and redMaPPer based on the SSDS data set. The two catalogues were independently constructed,  both aiming at very low false-detection rates. By comparing the two-catalogues against each other, the present paper investigates a number of practical issues critical for cluster studies and, therefore, goes much beyond the blind correlation analyses.
In particular, we performed an interactive screening  of the clusters found NOT to have either an X-ray or an optical counterpart, in order to disentangle possible technical detection problems from astrophysical biases and thus better understand the selection functions of the two samples. Among the questions we address, we cite:  What fraction of the non-matches can be ascribed to detection pipeline failures? Do the X-ray and optical detection pipelines miss any massive cluster? Do we find any optically rich cluster without X-ray gas beyond what is expected given scaling relations with log-normal scatter? To what extent is the optical sample contaminated by projection effects? How many X-ray clusters are missed because of the presence of a bright central AGN?  \\ 
 The paper is organised as follows. The next section summarises the properties of the 
X-CLASS and redMaPPer catalogues; Section 3 describes the adopted correlation procedures; Sections 4 and 5 scrutinise the correlation statistics for the optical to X-ray and X-ray-to-optical directions, respectively;  the results are discussed in Section 6 and last section draws the conclusions.
Throughout the article we assume the WMAP7 cosmology (\cite{komatsu11,larson11}).


\section{Catalogue overview}
The X-CLASS and redMaPPer galaxy cluster catalogues pertain to very different data types and detection methods and have already been published. In this section, we briefly summarise the properties of the two samples that are relevant for the present study.

\subsection{The X-ray catalogue}

\begin{table*}
\caption[]{X-ray source classification in terms of pipeline analysis variables. Contamination rates have been determined by means of extensive simulations}
\begin{center}
\begin{tabular}{lcccc}
\hline \hline \\
Source label & {\tt Extent} (1) & {\tt Ext-LH} (2) & {\tt PNT-LH} (3) & Contamination\\
\hline \\
C1 clusters & $> 5"$ & $>33$ & & $<5\%$ (4) \\
C2 clusters & $>5"$ & $ 15 < $ AND $ <33 $ & & $\sim 50 \%$ (4)\\

Clusters not analysed & & & NaN&  \\
(generally very nearby objects) & & & & \\
Clusters with strongly peaked core & $>30"$&  & $>60$& \\
(generally very nearby objects) & & & & \\
\hline \\
P1 point sources & $<3"$ & & $>30$ & $< 5\%$ (5)\\
\hline\\
W sources (weak) & & & $ >15$ & \\
M sources (marginal) & & & $<15$ & $\sim 50\%$ (6)  \\
\hline \hline
\label{XSRClabel}
\end{tabular}
\end{center}
\footnotesize
(1) Fitted core radius, assuming a $\beta =2/3$ profile after PSF deconvolution \\
(2) {\tt Extent\_Likelihood} \\
(3) {\tt Detection\_Likelihood}\\
(4) contamination by missclassified point-sources\\
(5) contamination by missclassified extended sources\\
(6) fraction of spurious sources
\end{table*}

\subsubsection{X-CLASS clusters}

The X-CLASS sample results from a serendipitous cluster search involving XMM archival observations performed until May 2010. Out of these, only observations at galactic latitudes higher than $|b|= 20$ deg were considered, and regions such as the Magellanic Clouds or the surroundings of bright nearby galaxies (e.g. M31) were excluded. Cluster detection was performed using the two-step XMM-LSS pipeline (Xamin), combining wavelet
multi-resolution analysis and maximum likelihood fits that make proper
use of Poisson statistics. The working radius of the pipeline was restricted to 13 arcmin (XMM has a total field of view of R=15 arcmin, but beyond R=13 arcmin, sensitivity and PSF are strongly degraded, therefore rendering cluster detection and characterisation unreliable beyond this radius).

\begin{figure}[t]
\begin{center}
\includegraphics[width=0.5\textwidth]{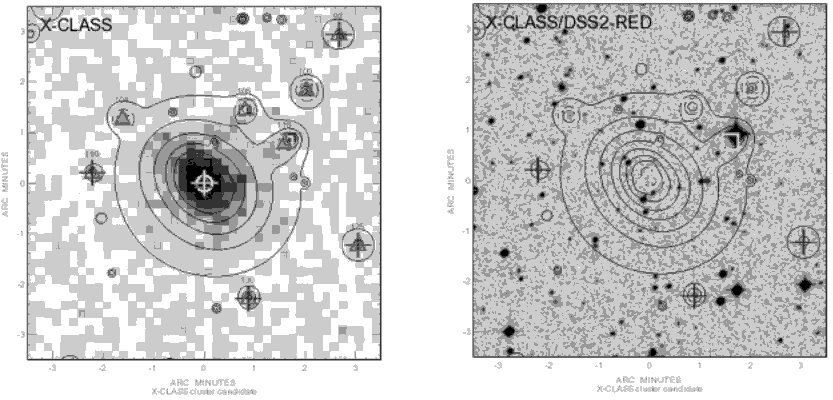} 
\caption{Example of a C1 cluster (X-CLASS 2305, $7'\times 7'$ image). \textit{Left panel}~: Wavelet-filtered contours superposed on the raw X-ray photon image. \textit{Right panel}~: DSS-II r-band image + wavelet contours. The X-ray contours are displayed for the [0.5-2] keV band. }
\label{clusterexample}
\end{center}
\end{figure}

The whole procedure has been evaluated by means of extensive image simulations (\cite{pacaud06}). This enabled us to create an {\sl
uncontaminated} (C1) cluster sample  by selecting sources in the [{\tt extent-extent\_likelihood}] output parameter space. From the simulations, we derived the probability for a cluster of given apparent size and flux to be detected as a C1 source. 
We emphasise again here that, unlike what has been commonly assumed so far, complete and uncontaminated cluster samples cannot be defined by a single flux limit, unless the limit is set very high compared to the survey sensitivity. Rather, cosmological cluster samples are limited by surface brightness and therefore must be selected in a two-dimensional parameter space. 
The X-CLASS catalogue (including the selection function) is described and published by \cite{clerc12b}. We stress below  some  features that are especially relevant for the present study.
\begin{itemize}
\item[•]  All processed XMM observations were cut to 10 ks of clean observing time on each of the three detectors. This enables us to ensure homogeneity  and, thus, ease calculation of the selection function (in this case, only background variations from pointing to pointing have to be considered). The 10 ks XMM exposures correspond to a point-source sensitivity of $\sim 5~10^{-15}$ \flux\  in the [0.5-2] keV band (80\% completeness limit, {\tt detection\_likelihood} > 15 and median background).
\item[•] XMM pointed observations of clusters were not excluded from our processing of the archive. This is a significant difference from other archive processings, such as the XMM Cluster Survey (XCS, \cite{mehrtens12}) and 2XMMi/SDSS Galaxy Cluster Survey (T13, \cite{takey13}). The main reason for this choice is that excluding some 200 clusters would significantly decrease the final sample. Furthermore, it would introduce a bias that is not a priori less than when including them, all the more so since one tends to propose XMM observations of the brightest clusters at any redshift  (see discussion in Sec. 3.5 of \cite{clerc12b})

\item[•] All sources flagged as C1 by Xamin were interactively screened by means of XMM/DSS overlays. The purpose of this procedure, which involved at least two different persons,  is twofold: (1) remove nearby galaxies, saturated point-sources, X-ray artefacts, and possible unresolved double sources that also appear as extended sources, and (2) provide an approximate distance indicator that depends on the existence of a conspicuous optical counterpart to the X-ray emission, namely: NEARBY ($z<0.3-0.4$) and DISTANT ($z>0.3-0.4$), where $z \sim 0.3-0.4$ corresponds to the POSS-II plate limit.
\item[•] On the basis of extensive simulations, the final C1 sample is estimated to have a degree of purity greater than 95\%. These clusters have a typical mass of $M_{500}\sim 5\times 10^{13} M_{\odot}$ at a redshift of 0.3 \citep{pacaud07},  where $M_{500}$ is the mass included inside a radius for an overdensity of 500 with respect to the critical density of the universe at the cluster redshift. Theoretical calculation of the mass limit as a function of redshift for this X-ray surface  brightness-selected cluster sample can be found in \citet{pierre11} (Fig. 2).
\item While only 420 high S/N clusters are published by \cite{clerc12b},  the present study makes use of the full X-CLASS 10ks catalogue containing  663 C1 clusters.     
\end{itemize}
The public X-CLASS sample is available at (\hyperref[http://xmm-lss.in2p3.fr:8080/l4sdb/]{http://xmm-lss.in2p3.fr:8080/l4sdb/}) and provides XMM/DSS overlay, as well as the details of the corresponding XMM observations. An example of an X-CLASS cluster is displayed in Fig. \ref{clusterexample}

\subsubsection{The other X-ray sources}
In addition to the C1 sample, we complement our optical to X-ray correlation analysis by considering all other sources detected in the 10 ks pointings used for the X-CLASS selection.
These sources can be classified as follows:
\begin{itemize}
\item[•] C2 sources constitute a second fainter cluster sample  so as to allow for $\sim$ 50\% contamination by misclassified point sources that can be cleaned up \textit{a posteriori} using X-ray/optical comparisons. 
\item[•] The Xamin pipeline was designed to ensure the best sensitivity for low surface brightness objects in Poisson regime. It is thus not adapted to characterising very nearby clusters that fill most of the XMM field of view (FOV), producing thousands of photons and leaving very little area for background estimates. Such sources are detected by the first pass of the pipeline but not analysed by the maximum likelihood module. A similar situation occurs for some weaker cluster sources located close to the border of the detection mask ($R_{off-axis}$ = 13 arcmin). These sources are characterised by a special flag on the {\tt Detection\_Likelihood} ({\tt Det\_LH}) parameter (NaN) and not included in the X-CLASS catalogue, meant to strictly contain only C1-type clusters. 
\item[•] Bright cluster sources that are strongly contaminated by a peaked central source (cool core or/and AGN).
\item[•] Unambiguous point-like sources constitute the P1 class and correspond to a S/N of at least 5 (Faccioli et al in prep.).
\item[•] The remaining sources are too faint (some 20 photons at most) to be characterised, given the XMM PSF and photon noise. They are split into two categories: (1) those significantly detected with a {\tt Det\_LH}  greater than 15 (weak sources, W-sources), and (2)  very marginal sources (M-sources) below this significance threshold. Our simulations showed that a large fraction of the latter are spurious. Our policy is to  only publish the $\tt Det\_LH>15$ sources \citep{chiappetti13}.
\end{itemize}
\noindent
The X-ray source classification is summarised in Table \ref{XSRClabel}. We stress that non-cluster sources constitute more than 90\% of the extragalactic X-ray source population at our sensitivity level. 

\subsection{The optical cluster sample}

The \redmapper\ is a new red-sequence photometric cluster finding algorithm which was recently applied
to the SDSS Data Release 8 \citep{dr8}. The algorithm and SDSS DR8 catalogue
is described in detail in \citet{rykoffetal14}.  A detailed comparison of \redmapper\ to other photometric
cluster finding algorithms is presented in \citet{rozo14}, which also includes a multi-wavelength study of the
performance of \redmapper\ in the SDSS.  A comparison of the \redmapper\ catalogue to the Planck SZ catalogue \citep{pXXIX} is presented in \citet{rozoetal14}.

Briefly, \redmapper\ models the red sequence of galaxy clusters as having Gaussian
scatter along a mean colour--magnitude relation. Both the mean colour--magnitude
relation and the scatter are parameterised via spline interpolation, with the free parameters
being the value at the nodes.  These values are iteratively self-trained by leveraging
both SDSS photometry and spectroscopy. We photometrically identify cluster galaxies
about galaxies with spectroscopic redshift and assign these photometric members to the cluster
redshift, allowing us to better calibrate the red sequence at faint magnitudes.  
The associated spectroscopic requirements for the above training are minimal, and are 
easily satisfied by existing SDSS spectroscopy.

Once the red-sequence model has been trained, the algorithm attempts to grow a galaxy
cluster centred on each SDSS photometric galaxy.  The galaxies are first rank-ordered
according to their likelihood of being a central galaxy.   We then proceed to estimate the
richness about the top central galaxy candidate.  The richness estimate $\lambda$ measures
the total number of red-sequence galaxies brighter than $0.2L_*$ within a cluster radius $R(\lambda)$,
selected to optimize the S/N of the measurement.  The richness is defined by
\begin{equation}
\lambda = \sum p_i
\end{equation}
where $p_i$ is the probability of galaxy $i$ of being a cluster galaxy, as estimated based 
on the red-sequence model calibrated above and the mean background of non-cluster galaxies,
estimated by computing the mean galaxy density across the entire SDSS DR8 footprint. 

Once a galaxy cluster has been identified ($\lambda\geq 5$, where
$\lambda$ is the number of red-sequence galaxies hosted by the cluster), 
the algorithm iteratively determines a photometric redshift
based on the calibrated red-sequence model, and re-centres the clusters about 
the best cluster centre\footnote{redMaPPer selects the central galaxies using an iterative matched filter algorithm.  In the initial run, the cluster centre is set to the cluster’s BCG.  These centres are used to estimate the filter describing the central galaxy luminosity and central galaxy density of the cluster.  These filters are then used in a new run of the cluster finding, and the procedure is iterated.  In the final catalogue, the central galaxy in a cluster is the same as the BCG only ~80\% of the time.}, as gauged from the photometric data.   
The final catalogue is then trimmed
further to a richness limit of $\lambda \ge 20$ for $z\leq 0.35$.  Above this redshift, the catalogue
becomes ``flux'' limited owing to the SDSS survey depth, and the richness limit increases
rapidly with redshift.  Roughly speaking, richness measurements are reliable out to
$z=0.5$.  For $z\in [0.5,0.6]$, clusters can be detected, but their richness measurements
become very noisy.  When run on SDSS data, 
automated cluster finding is not really feasible with the \redmapper\ algorithm above redshift $z=0.6$.
We note that the red-sequence model is trained over the redshift range $z \in [0.05,0.6]$.  Consequently, 
clusters near the redshift edges can have unreliable redshifts: robust photometric redshift
performance is expected to be limited to $z\in [0.08,0.55]$. \\
The photometric sensitivity is roughly but not exactly constant over the entire survey area.  Variations in depth as a function of position exist, but their dependences are not included in \redmapper\ v5.2.  Because SDSS is roughly uniform, however, the differences induced by these variations are small.  As for the masked area, the \redmapper\ mask is defined such that no cluster is masked by more than 20\% by the BOSS galaxy mask.\\
The \redmapper\ redshift detection range is $z \in$ [0.05,0.6].  The upper redshift limit is driven by the depth of the SDSS: the galaxies in clusters at higher redshifts are not detectable in the SDSS.  At low redshifts, the cluster selection is limited by the relative lack of rich galaxy clusters used for photometric calibration (due to small volume) and the fact that the photometry of very bright low redshift galaxies from the automated SDSS pipeline is often compromised (e.g. many galaxies are saturated, and therefore not present in the photometric catalogue).

\begin{figure}[t]
\begin{center}
\includegraphics[width=0.35\textwidth]{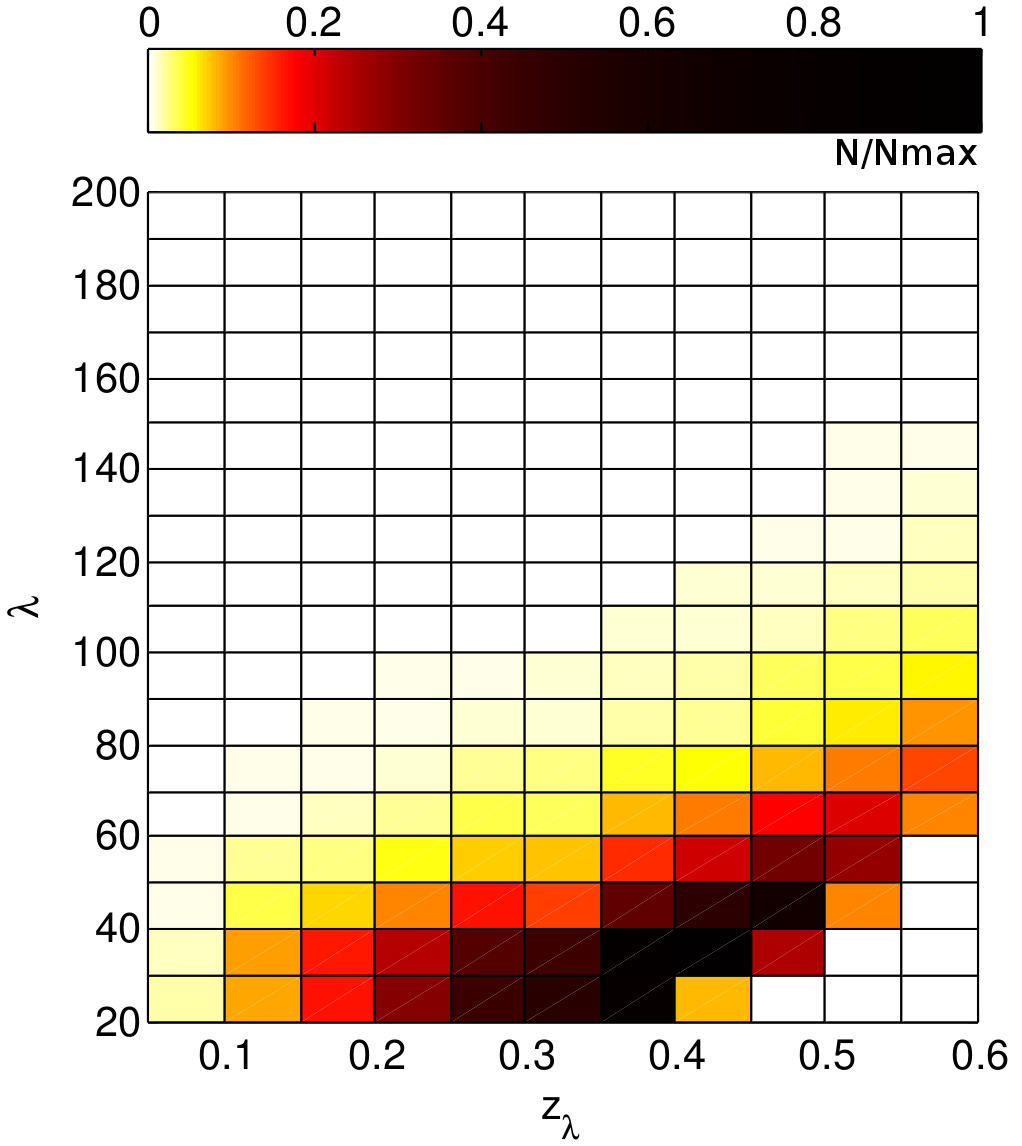} 
\caption{
\textit{Redshift-richness} distribution ($z_\lambda-\lambda$) of the entire redMaPPer catalogue (26,138 clusters). Colour coding is normalised by the highest pixel value of the diagram (Nmax).}
\label{redMaPPerprop}
\end{center}
\end{figure}

The \textit{redshift-richness} distribution of the sample is shown in Fig. \ref{redMaPPerprop}.

\begin{figure}[t]
\begin{center}
\includegraphics[width=0.35\textwidth]{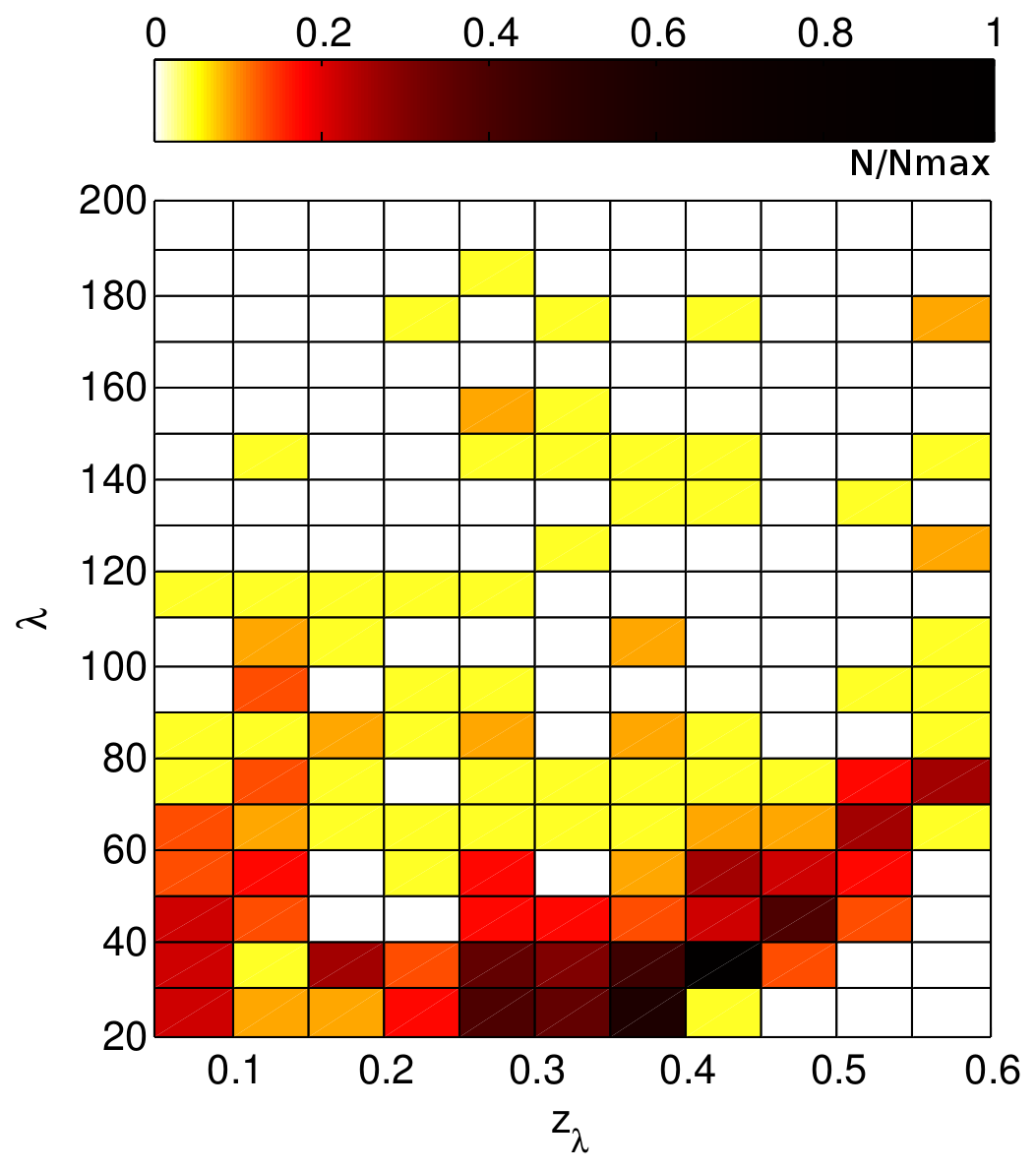} 
\caption{\textit{Redshift-richness} distribution ($z_\lambda-\lambda$) of the 270 redMaPPer clusters falling on X-CLASS pointings. Same colour coding as in Fig.\ref {redMaPPerprop}. }
\label{RedXprop}
\end{center}
\end{figure}

\section{Matching the optical and X-ray cluster samples}

\subsection{Optimisation of the matching radius}
When matching clusters in position, one needs to define the aperture used for the positional matching. This is an important and not  obvious choice. It must be large enough to account for the positional uncertainties of both catalogues.  However, its value should be limited by the condition that the number of chance associations will be kept to an acceptable level, when the optical sample is further correlated with the entire X-ray catalogue (i.e. including the numerous point-like or unresolved detections). Basically, there are two possibilities: use either a fixed angular scale or a physical scale, none of which is perfect, as discussed below.\\
The XC1 and redMaPPer  densities are $\sim$ 5/\dd\ (for a uniform X-ray coverage, $z_{lim}<1.5$)  and 2.5/\dd\ ($\lambda > 20$, $z_{lim}<0.6$), respectively.  For 10 ks exposures, the XMM source density above a detection likelihood of 15 is $\sim$ 400 per square degrees in the soft band  (\cite{chiappetti13}), and this corresponds to a mean separation between sources of 3 arcmin. Our experience with the optical follow-up of XMM-LSS clusters (\cite{pacaud07}) shows that, for a very large fraction of the C1 population, the X-ray centroid calculated by the pipeline matches the position of the cD galaxy better than 1/2 arcmin . Cases where larger offsets are observed ($\sim$ 10\%) correspond to obvious mergers for which the X-ray emission shows multiple maxima or is very flat; such merger situations also affect the determination of the optical centre. Optical centring is likewise prone to inherent uncertainties like, in the case of the redMaPPer algorithm, the wrong identification of the cD galaxy, which may put the calculated centre up to a few arcmin of its actual position.    \\
We have considered the possibility favouring a fixed physical scale, such as half of the cluster virial radius. Assuming a mean half-viral radius of 750 kpc (4.5 keV clusters, \cite{pratt09}) yields angular scales of 7$\arcmin$-1.9$\arcmin$ for the redshfit range of interest; such a radius will thus yield on average some 5.4 to 1.2 chance coincidences for the correlation with the point-source catalogue. Correlation lengths significantly larger than the one-arcmin scale will further hit increasingly significant "no man's land" regions, when looking for counterparts of redMaPPer clusters that fall close to the border of the XMM FOV (R = 13 arcmin). In addition, the $\approx 1\%$ redshift failure rate in redMaPPer \cite{rozo14} will compromise the use of fixed metric apertures for these systems.\\
Given these practical limitations, we  set a fix angular search radius of 1 arcmin. 
Between $0.1<z<0.6$ - where most of the clusters of interest for the present study lay - this angular scale spans 110$-$400 kpc.

\subsection{Overlap between the two catalogues} 

Any correlation study requires careful determination of the region of overlap between the catalogues.  Here, the exercise is complicated by the fact that the X-ray coverage is sparse with, moreover, some overlap between the individual XMM observations. Similarly, the  sky distribution of the redMaPPer clusters is affected by the BOSS masking. 
To ease the task, we have thus defined two samples with optimal overlap: one (OPT$\rightarrow$X) that will be used for studying the X-ray counterparts of the optical clusters and a second one (X$\rightarrow$OPT) for studying the optical properties of the X-ray clusters.

\subsubsection{Sample OPT$\rightarrow$X}

We first identified all XMM X-CLASS observations containing at least one redMaPPer position. Our search radius  was restricted to 12 arcmin from the centre of each X-ray observation to match the working radius of the X-ray pipeline (13 arcmin) when allowing for 1 arcmin source positional errors (see Sec 3.2). 
In total, we have identified 223 X-CLASS pointings  containing 270 redMaPPer cluster positions. The corresponding X-ray area is $\sim 27$ \dd , once the overlap between the XMM pointings is removed. The properties of these objects are presented in Fig. \ref{RedXprop};  differences from the complete sample (Fig. \ref{redMaPPerprop}), especially in the first two redshift bins  can be ascribed to the presence of XMM pointed observations on some clusters of interest (see discussion in 5.1).

\begin{figure*}
\begin{center}
\includegraphics[width=0.9\textwidth]{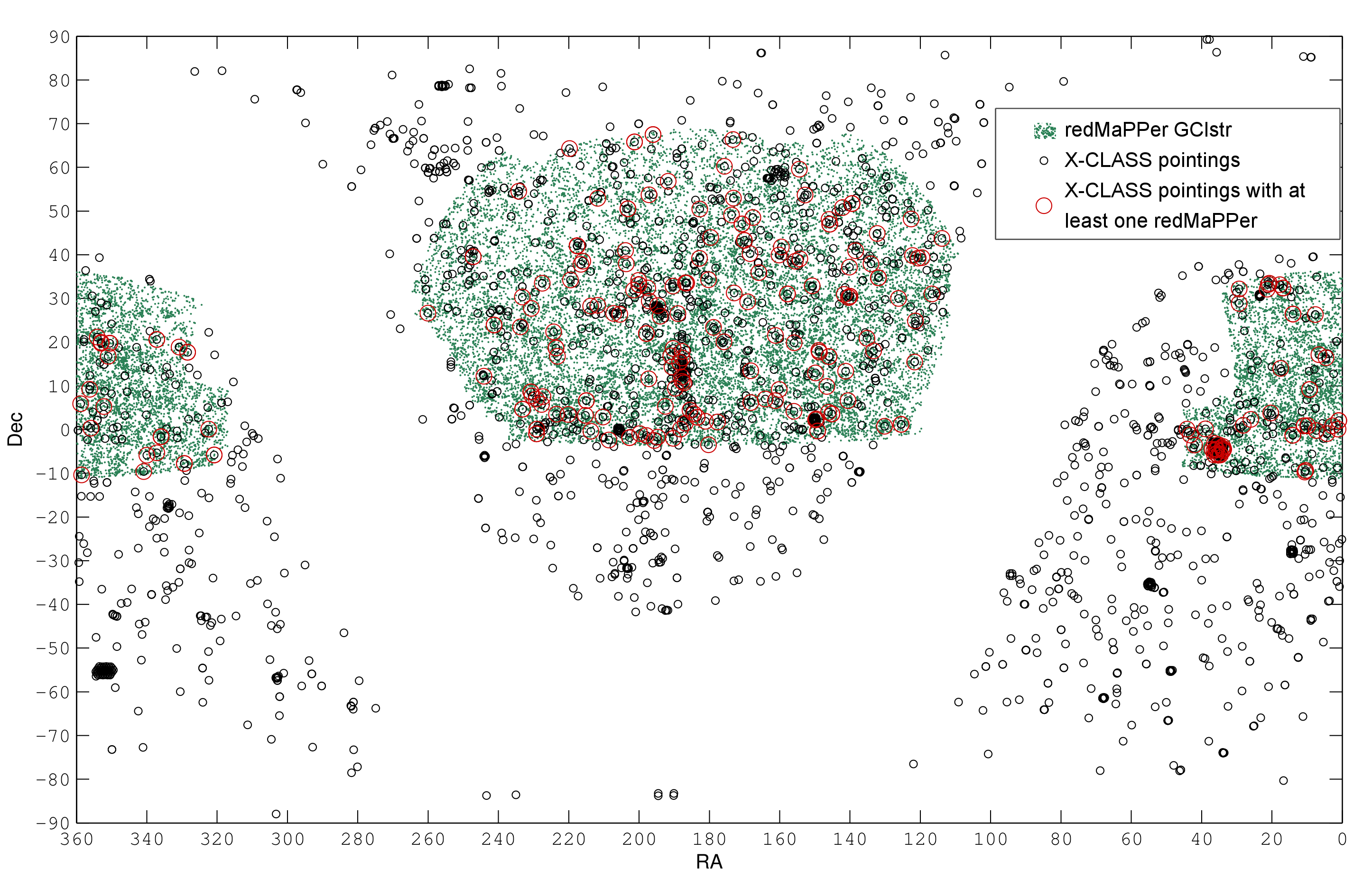} 
\caption{Sky coverage of the redMaPPer and X-CLASS catalogues. The green dots stand for the redMaPPer galaxy clusters (26,138 objects, $\lambda > 20$). The black circles indicate the XMM-Newton pointings pertaining to the X-CLASS catalogue (2409 observations); the size of the XMM FOV (30$\arcmin$) is exaggerated in the figure; The red circles flag the X-CLASS pointings that contain at least one redMaPPer cluster within $12 \arcmin$ (223 observations). }
\label{sky1}
\end{center}
\end{figure*}

The sky distribution of the redMaPPer catalogue and XMM pointings is displayed in Fig. \ref{sky1}.  In the following, we name `XC1' as the sub-sample of C1 X-CLASS clusters falling onto the region covered by the redMaPPer catalogue and XC1$^{+}$, the XC1 sample supplemented by the clusters  manually recovered ({\tt detection\_likelihood} = NaN, 29 objects, see Sec. 4.2 item 3).

\subsubsection{Sample X$\rightarrow$OPT}

\begin{figure*}
\begin{center}
\includegraphics[width=0.9\textwidth, height=0.3\textheight]{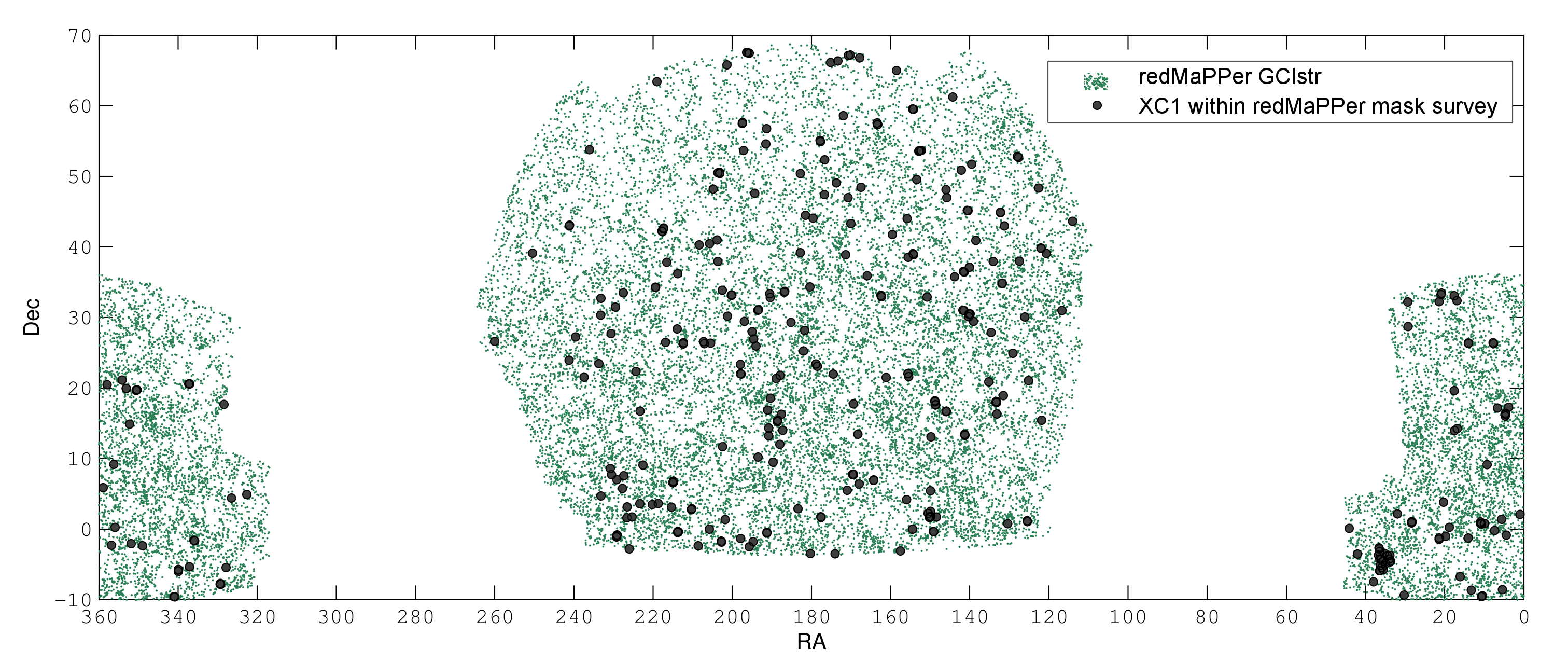} 
\caption{Overall area encompassed by the redMaPPer catalogue (BOSS survey) is mapped by the green dot distribution as in Fig. \ref{sky1}. The black points show the 355 XC1$^{+}$ clusters  that have passed the redMaPPer area masking (exclusion of `bad' fields and of the vicinity of bright stars).}
\label{sky2}
\end{center}
\end{figure*}

The second step was to identify which X-ray clusters have a potential redMaPPer counterpart.
For this, we made use of the masks defined for constructing the redMaPPer catalogue: footprint of the BOSS survey minus the regions flagged as a `bad field' or excluded areas in the vicinity of bright stars. In total,  355 XC1$^{+}$ were found to pass the mask selection and constitute the X$\rightarrow$OPT sample.  The optical properties of this sample are presented in Sect. \ref{XvsO} and its sky distribution shown in Fig. \ref{sky2}.

\begin{figure*}[t]
\begin{center}
\includegraphics[width=0.9\textwidth]{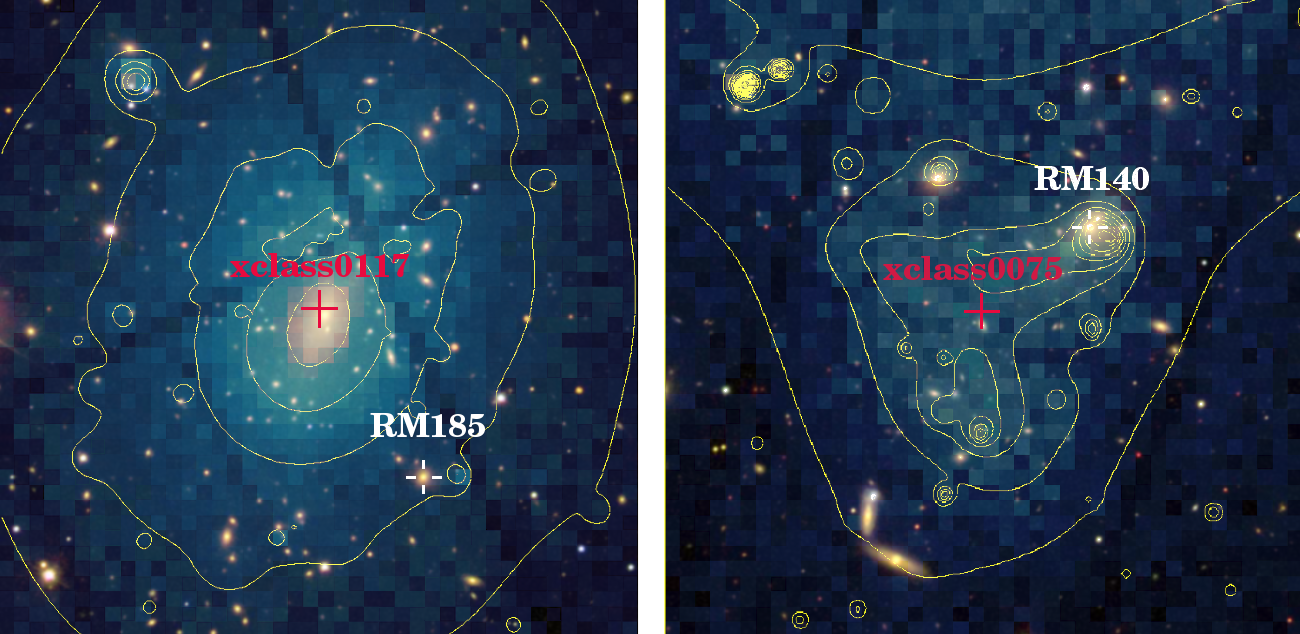} 
\caption{Two problematic matching cases, which were a posteriori recovered by visual inspection. The SDSS colour $7' x 7'$ image is overlaid on the raw X-ray photon image + contours (the X-ray image is not corrected for detector cosmetic); the optical and X-ray centres are marked by a white and a red cross, respectively. \textit{Left}~: Wrong optical centre; the offset between the redMaPPer and X-ray centres is $2.2'$.  \textit{Right}: Complex nearby structure; the offset between  the two centres is  $1.7'$.}
\label{complexmatch}
\end{center}
\end{figure*}

\subsection{A posteriori checks}

We  have visually inspected all clusters for which no or several counterparts were found, in order to correct our correlation statistics a posteriori for any shortcoming of the X-ray or optical pipelines. Two examples of a complex matching configuration solved by visual inspection are presented in Fig. \ref{complexmatch}. We finally  checked whether the 1 arcmin correlation disk around each of the 355 XC1$^{+}$ positions was significantly affected by the optical masking: no masking greater than 30\% was found.\\

\section{The X-ray counterpart of the redMaPPer clusters}
\label{OvsX}

For this study we use the OPT$\rightarrow$X sample (270 redMaPPer clusters) as defined in Sec. 3.1.1

\subsection{Statistics}
\begin{figure}[t]
\begin{center}
\includegraphics[width=0.43\textwidth]{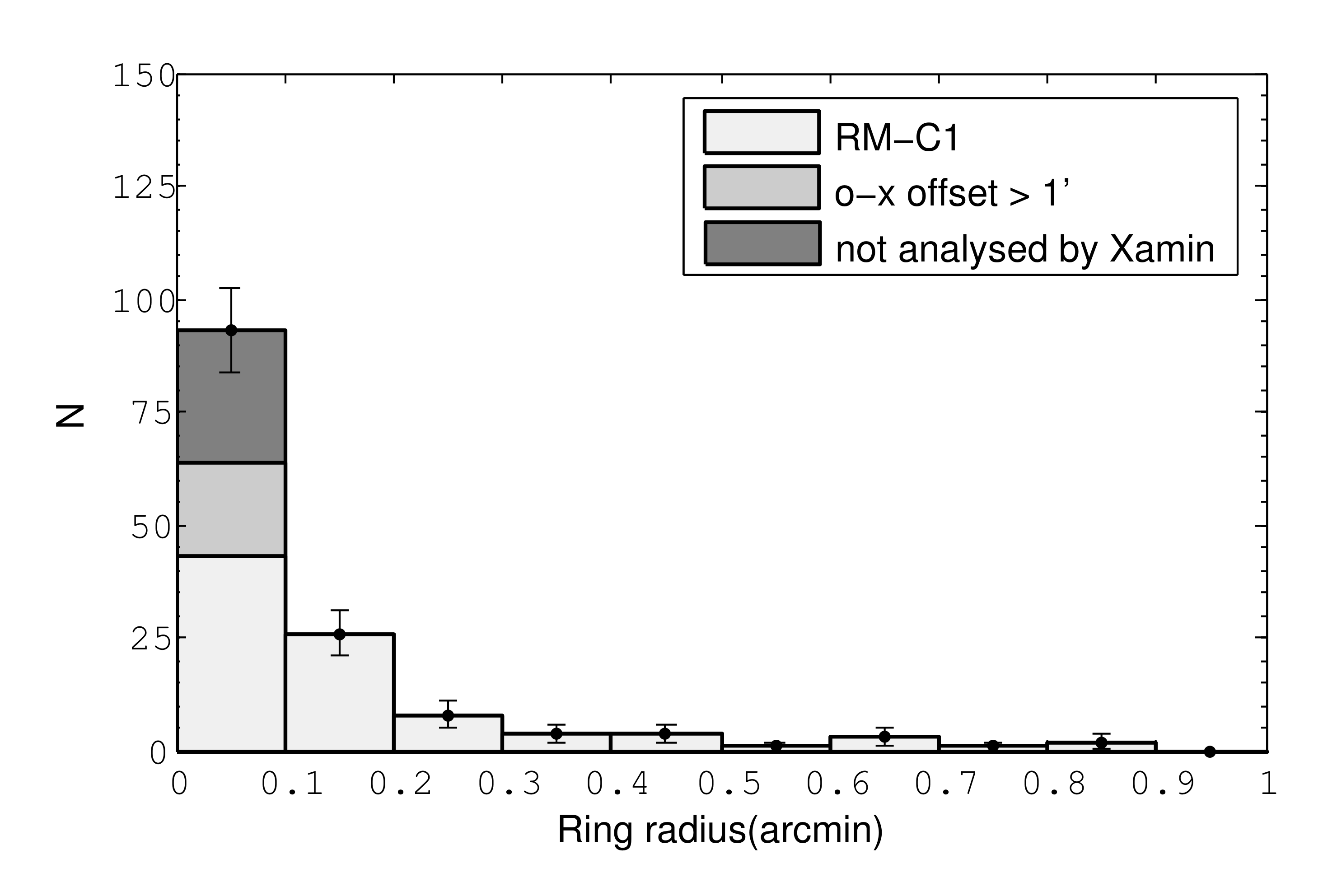} 
\caption{Offsets between the redMaPPer  and the XC1 cluster positions (142 objects in total). Clusters missed because of failures of the X-ray or optical pipelines but recovered after visual inspection of the data are added in the first bin.}
\label{offset-cluster}
\end{center}
\end{figure}

\begin{figure}[t]
\begin{center}
\includegraphics[width=0.35\textwidth]{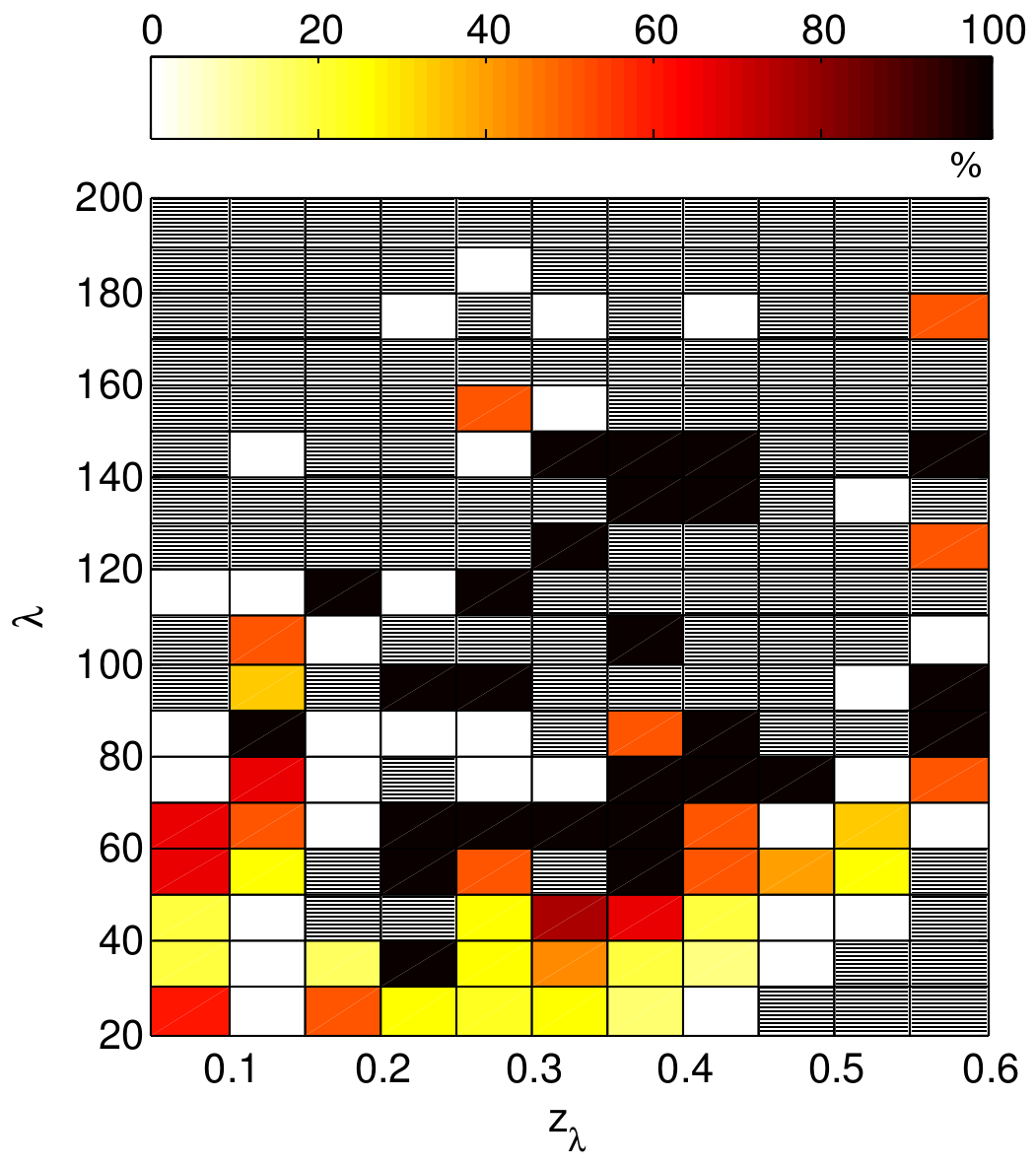} 
\caption{Distribution of the redMaPPer clusters detected as XC1, as a function of richness and redshift. In total, 92 matches with C1 X-ray clusters for 270 input redMaPPers are found within a radius of 1$\arcmin$. A posteriori recovered correlations are not included in this plot (see Fig. \ref{XC1asredMaPPerprop} for a complete census). Colour coding gives the percentage of redMaPPer detected as C1 in each diagram pixel; greyed pixels stand for \textit{redshift-richness} combinations not present in the input OPT$\rightarrow$X catalogue (Fig. \ref{RedXprop}) }
\label{RedXprop2}
\end{center}
\end{figure}

We first matched the redMaPPer clusters  with the XC1 clusters within a radius of  1$\arcmin$. We found 92 ($\sim34\%$) matches. Figure \ref{offset-cluster} shows the distribution of the corresponding optical-X-ray separations:  92\% of the matches occur for distances less than 0.5 arcmin, and their optical characteristics  are shown in Fig.\ref{RedXprop2}. This distribution is expected to largely reflect the centring offsets between X-ray centres and the true central galaxies of clusters and/or positional uncertainties in the X-ray centre.

\subsection{Inventory of the redMaPPer clusters not detected as C1}
Each of the 178 redMaPPer clusters not matched with an XC1 within 1$\arcmin$ was then examined by eye on the basis of X-ray/optical overlays. We further correlated these clusters with the complete X-ray source list  as described in Table \ref{XSRClabel}.   
We review below the outcome of this analysis case by case. (The percentages are expressed as a function of the OPT$\rightarrow$X cardinal i.e. 270 objects.)

\begin{figure}[t]
\begin{center}
\includegraphics[width=0.43\textwidth]{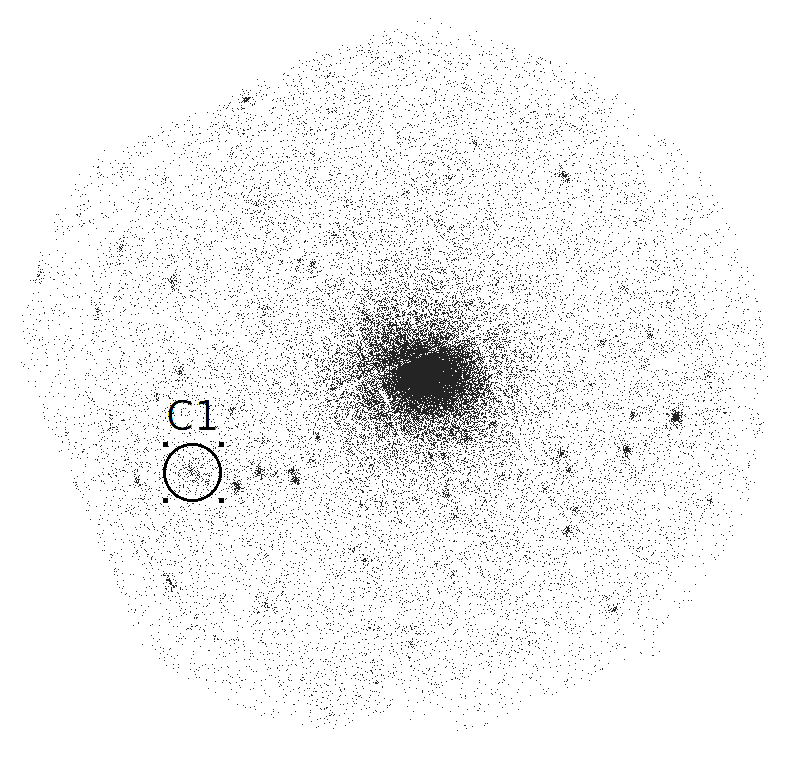} 
\caption{Example of a very bright cluster detected but not analysed by the X-ray pipeline (Abell 773). The faint extended source to the east is detected as a C1 cluster. The full XMM FOV is presented}
\label{bigmiss}
\end{center}
\end{figure}

\begin{figure}[h]
\begin{center}
\includegraphics[width=0.5\textwidth]{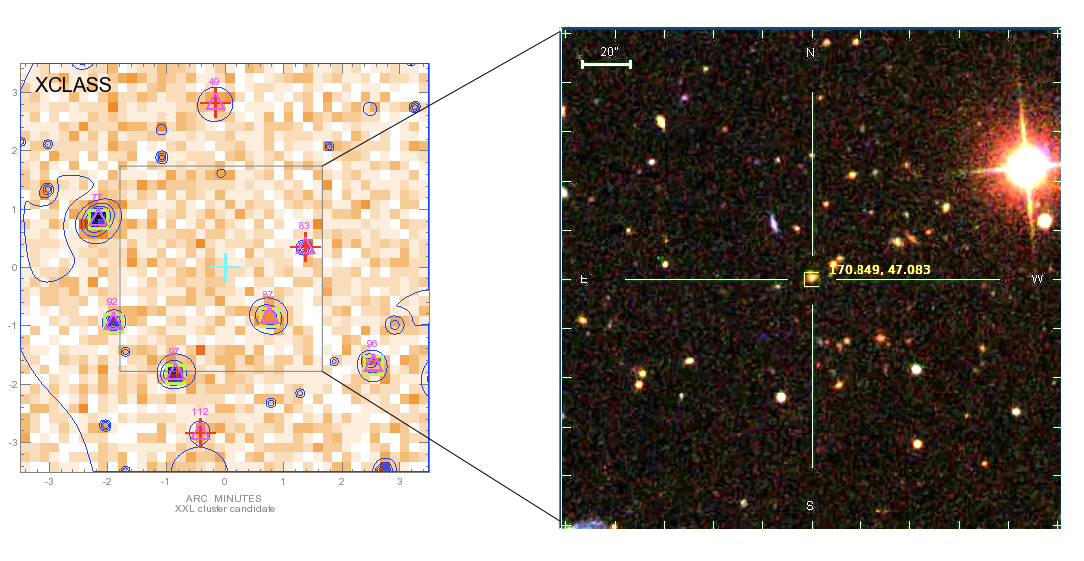}  
\caption{redMaPPer cluster 29357 ($z_{lam} =0.48$ and {\em Richness} = 44.15). \textit{Left panel}: $7\arcmin\times7\arcmin$ [0.5-2] keV photon image; green squares stand for point sources and red crosses for detections below the significance level ({\tt Det\_LH} < 15); \textit{Right panel}: corresponding SDSS DR9 color $3\arcmin\times3\arcmin$ image.}
\label{noX}
\end{center}
\end{figure}

\begin{description}
\item[1.    ] {$\sim 5\%$} (17) of the redMaPPers are found to have a C2 counterpart within 1$\arcmin$. Most of them have a richness $\lambda < 40$ and a redshift $0.1 < z < 0.5$ with a median value of $\sim 0.38$. 

\item[2.    ] {$\sim 2\%$} (7) are associated with an extended emission plus strong central peak. 

\item[3.] {$\sim 11\%$} (29) are associated with detected X-ray sources but not analysed by the Xamin pipeline (mostly very nearby objects filling a large fraction of the detector). Such non-matches were manually recovered and a posteriori added to the XC1 sample to constitute the XC1$^{+}$ sample. An example is shown in Fig. \ref{bigmiss}.

\item[4.    ] {$\sim 7\%$} (19) are associated with a P1 point source  within 1$\arcmin$.

\item[5.    ] {$\sim 10\%$} (27) are associated with a W source  within 1$\arcmin$.

\item[6.    ] For completeness, we mention that {$\sim 9\%$} (23) of the redMaPPer are associated with an M source  within 1$\arcmin$. We recall that this last category is defined by a detection likelihood lower than 15, and it contains about 50\% X-ray detections.

\item[7.    ] \textbf{$\sim 8\%$} (21) of the redMaPPers show a large offset between the optical and C1 X-ray positions, i.e. more than $1\arcmin-3\arcmin$. The eye inspection showed that for many of these objects, the cD was misidentified. These matches were recovered after visual inspection.

\item[8.    ] \textbf{$\sim 13\%$} (35) of the redMaPPer are not associated with any X-ray detection within 1$\arcmin$. An example is shown in Fig.\ref{noX}. 
\end{description}

\begin{figure}[t]
\begin{center}
\includegraphics[width=0.43\textwidth]{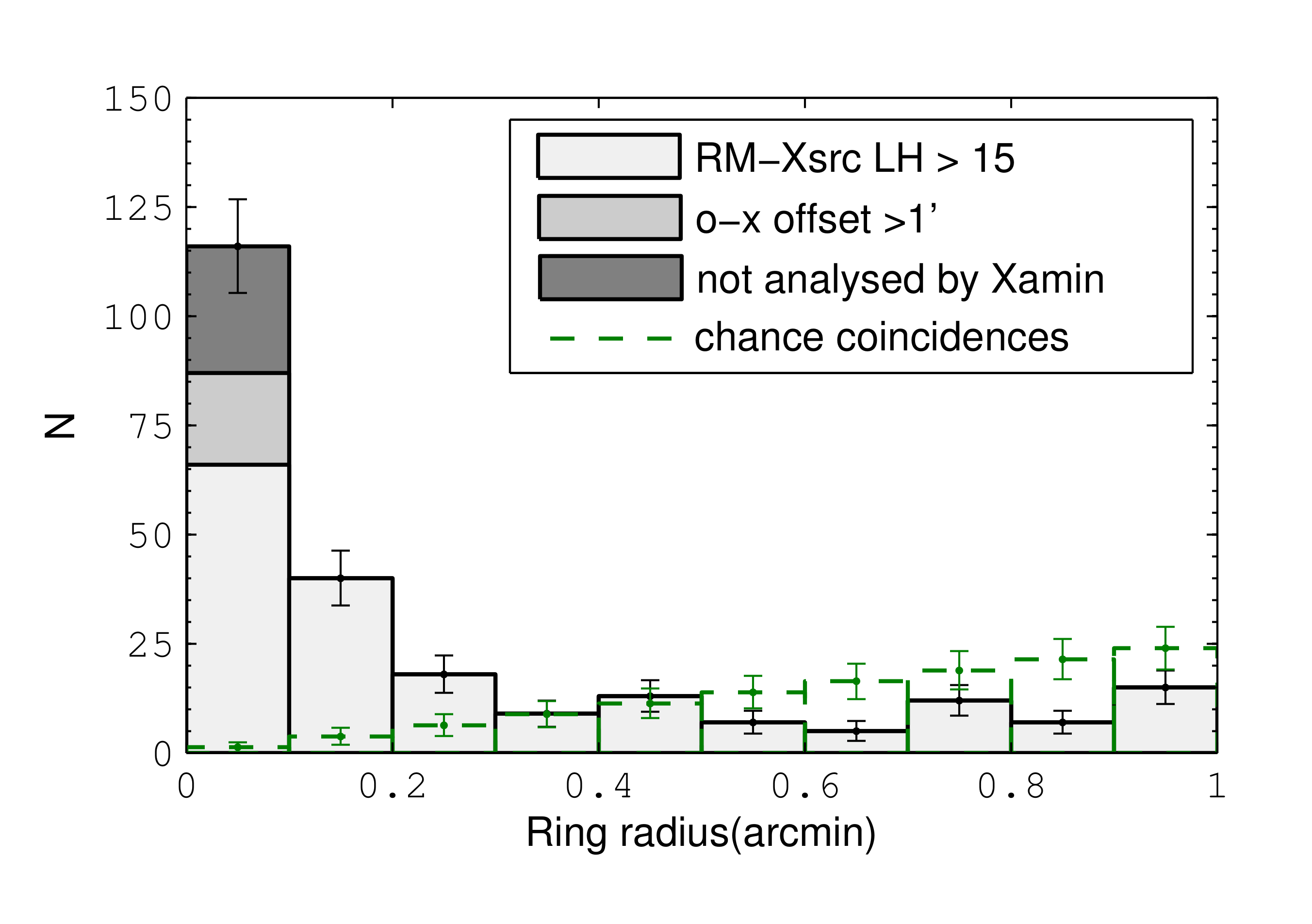} 
\caption{Offsets between the redMaPPer  and   X-ray  source positions (of any type, {\tt Det\_LH} > 15) superposed on the calculated distribution of chance coincidences (computed from the entire population with {\tt Det\_LH}>15). The breakdown of the 212 matched objects is as follows: light grey histogram: C1 plus categories 1, 2, 4, 5 of Sec. 4.2; medium grey histogram: category 7; dark grey histogram: category 3. Clusters missed because of failures of the X-ray or optical pipelines  (categories 3 and 7)  but recovered after visual inspection  are added in the first bin. }
\label{offset-point}
\end{center}
\end{figure}
\noindent
Figure \ref{offset-point} shows the distribution of the distances between the optical and X-ray source centres. \
After correcting for the large offsets  and not analysed  clusters, 53\% of the redMaPPer have a C1-type counterpart (C1 + categories 3, 7).
In total  $79\%$ ( C1 plus categories 1, 2, 3, 4, 5, 7) of the redMaPPer were found to coincide with an X-ray source (not considering the M sources);
The correlation break-down is illustrated in Fig. \ref{summary-match}.
As easily understandable, a large fraction of the clusters not analysed by Xamin lie in the upper part of the diagram (rich objects, hence expected to be X-ray bright). Undetected redMaPPer clusters lie mostly along the redMaPPer sensitivity limit. A summary of the detection statistics is presented in Table \ref{finalstat}.
  We noted that, below $z<0.25$, offsets from P1 or weak sources are all smaller than 0.15'; this may suggest that these matches are mostly due to the emission of the central galaxy.\\
We finally mention that we also performed the OPT$-$>XC1$^{+}$ correlation restricting the redMaPPer catalogue to its low-redshift part. Within 1 arcmin, the matching rates are 45\% and 56\% for the $z<0.5$ and $z<0.3$ sub-samples, respectively, to be compared to 45\% for the full sample, which contains twice as many clusters as when limited to $z<0.3$. Correcting for positional offsets, these fractions are increased to 53\% and 70\%, respectively.

\subsection{Stacking analysis of the redMaPPer with no X-ray counterpart}

In this section, we consider the cumulative X-ray signal of the redMaPPer clusters that have no XMM  counterpart at all (within our sensitivity limits) and, for comparison, those having a marginal counterpart (M-sources) i.e. objects falling in above-categories 8 and 6, respectively (black and orange stars in Fig. \ref{summary-match}). 
We split the analysis into two subsamples: redMaPPer positions falling within an off-axis angle of $9'$ or between $9'-12'$, knowing that beyond $9'$, PSF blurring and vignetting strongly reduce the  detection limit.  For this study, we consider the X-ray images that were used for the C1 detection, i.e. cut to an exposure time of 10 ks.  \\

The stacking analysis of XMM data is a particularly challenging task given the topology of the focal plane (combination of three telescopes, gaps between the CCDs)  and should account for the PSF and sensitivity variations as a function of off-axis distance. Furthermore, the mean background is subject not only to local cosmic variations (on a few arcmin scale) but also to the observing conditions inherent to the revolution in question (solar activity) and depends on the position of the spacecraft on the orbit at the observing date. The background consists of two components: the cosmic background, which is subject to vignetting and the particle background uniformly hitting the detector\footnote{http://xmm2.esac.esa.int/external/xmm\_sw\_cal/background/}. To this, must be added the fact that the signal is Poissonian in our case: as can be appreciated in Fig. \ref{noX} many of the 6 arcsec pixels are still empty after 10 ks.\\

\begin{figure}[h]
\centerline{
\includegraphics[width=0.5\textwidth]{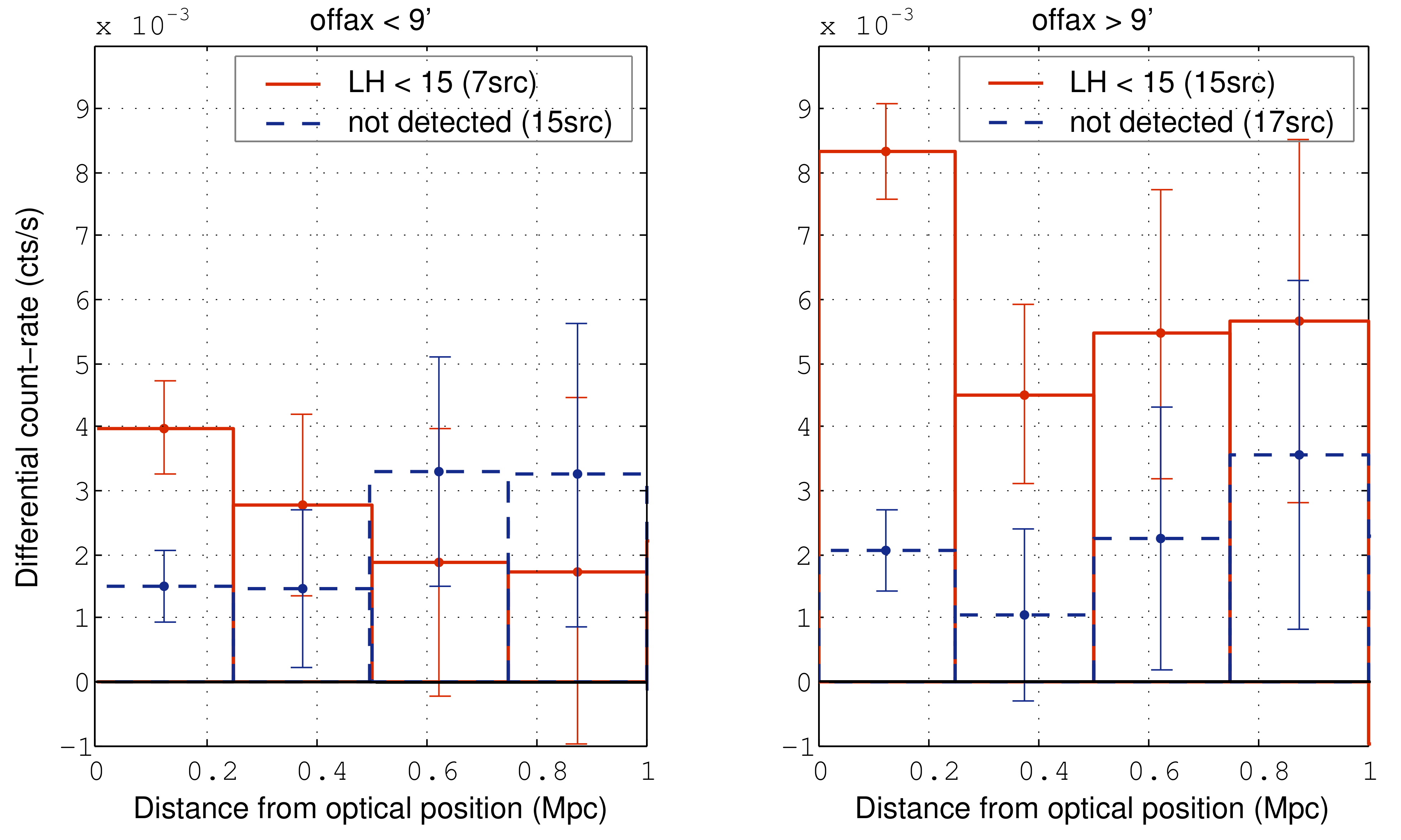}
}
\caption{Stacking analysis of the  redMaPPer clusters not associated with any  X-ray detection or coinciding with a marginal source ({\tt detection\_likelihood < 15)}. \textit{Left panel:} redMaPPer positions located at an off-axis radius $< 9\arcmin$. \textit{Right panel:} redMaPPer off-axis positions between $  9-12 \arcmin$.}
\label{noXreal}
\end{figure}

\begin{figure}[h]
\centerline{ 
\includegraphics[width=0.5\textwidth]{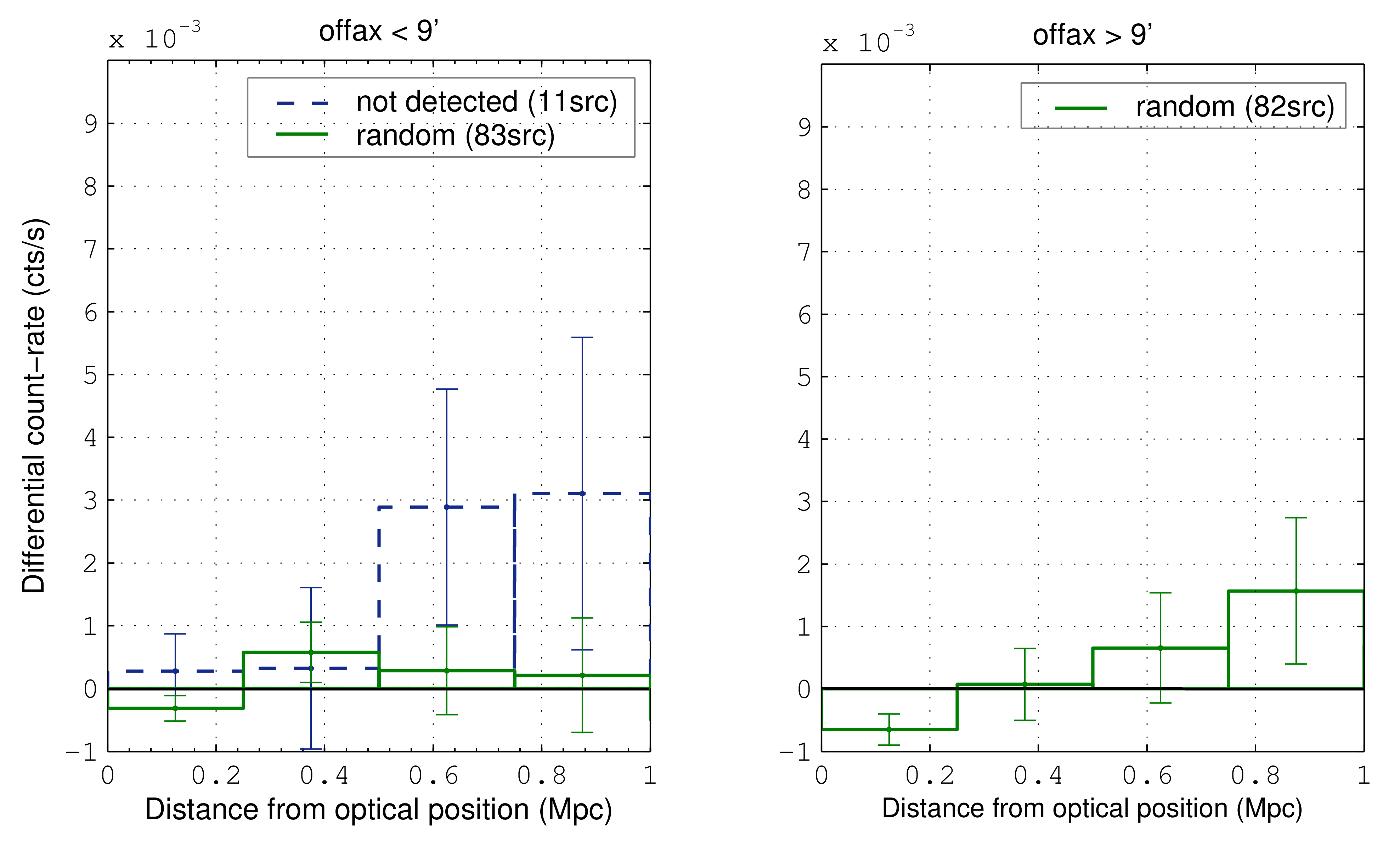}
} 
\caption{Stacking analysis around random positions, not associated with any  X-ray detection within 1 arcmin. \textit{Left panel:} offaxis$ < 9\arcmin$, \textit{Right panel:} offaxis$ > 9\arcmin$. Overlaid on the left panel is the signal obtained when removing four redMaPPer positions not associated with any detected source but for which some intensity enhancement is conspicuous on the X-ray pixel images.}
\label{noXrand}
\end{figure}

To circumvent these practical difficulties, we have adopted a procedure that does not stack the photon maps, but rather the putative X-ray profiles. Profiles were calculated from a growth curve analysis, centred on the optical position, so that they estimated the background in the vicinity of the putative cluster within a  $150\arcsec-500\arcsec$ annulus (i.e. $1-3.3$ Mpc at $z=0.6$, the maximum redshift of the sample); the principle of the analysis  is described by \cite{clerc12b}. Here, profiles were computed for a physical bin size of 250 kpc, and given the boundaries of the background annulus, all profiles were stopped at 1 Mpc. Error bars for each bin were scaled to the mean statistical fluctuations as determined in the background and source annuli. The individual cluster profiles (which may show some negative values) were then averaged for both categories, each data point being weighted by the inverse of the square of its error bar.\\
We have further defined a control sample consisting of random positions thrown in the XMM pointings pertaining to the sub-sample. We kept only positions that do not coincide with any X-ray source within a radius of one  arcmin. In total, 165 positions (out of the 1000 simulated ones) were retained and subsequently assigned a redshift so as to match the redshift distribution of the sub-sample. Results of the stacking analysis for the science and random samples are displayed in Figs. \ref{noXreal} and \ref{noXrand}, respectively. We discarded three clusters out of the 35 in Category 8 of Sect. 4.2 since they appeared to be located in the vicinity of bright nearby clusters, hence not allowing a reliable background determination. On the left-hand side of Fig. \ref{noXrand}, we overlaid the signal obtained when removing four Category-8 clusters,
visually found to coincide with some X-ray enhancement: the remaining signal is compatible with 0. One example of removed clusters is displayed in Fig. \ref{bordercase}.

\section{The optical counterpart of the X-CLASS clusters} 
\label{XvsO}
In this section, we present the cross-matching between the X-CLASS and redMaPPer samples and use the X$\rightarrow$OPT sample.  
\subsection{Statistics}
\begin{figure}[t]
\begin{center}
\includegraphics[width=0.35\textwidth]{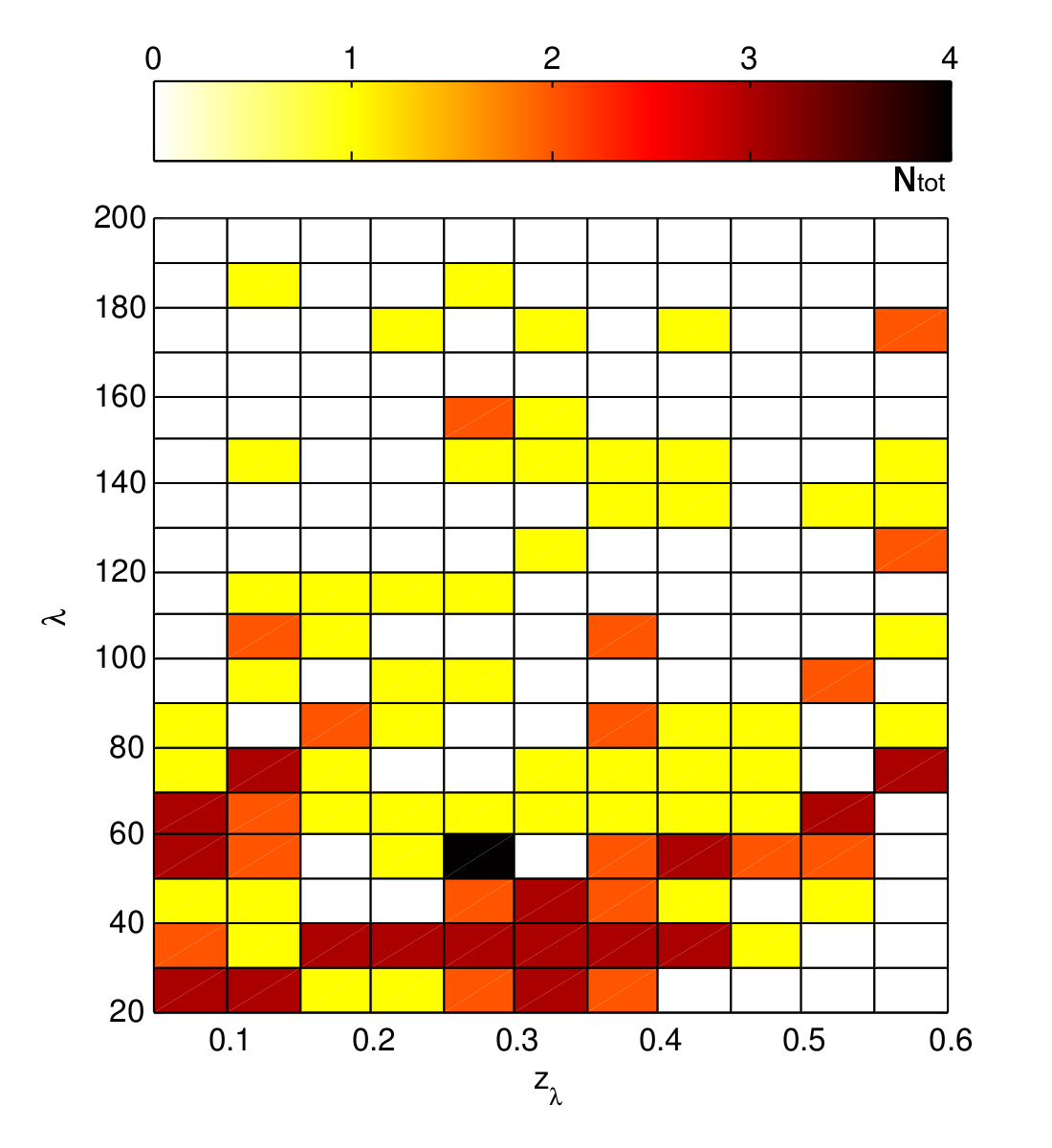} 
\caption{\textit{Redshift-richness} distribution of the 144 XC1$^{+}$ clusters detected as redMaPPer. Colour coding gives the number of objects in each diagram pixel. }
\label{XC1asredMaPPerprop}
\end{center}
\end{figure}

\begin{figure}[t]
\begin{center}
\includegraphics[width=0.35\textwidth]{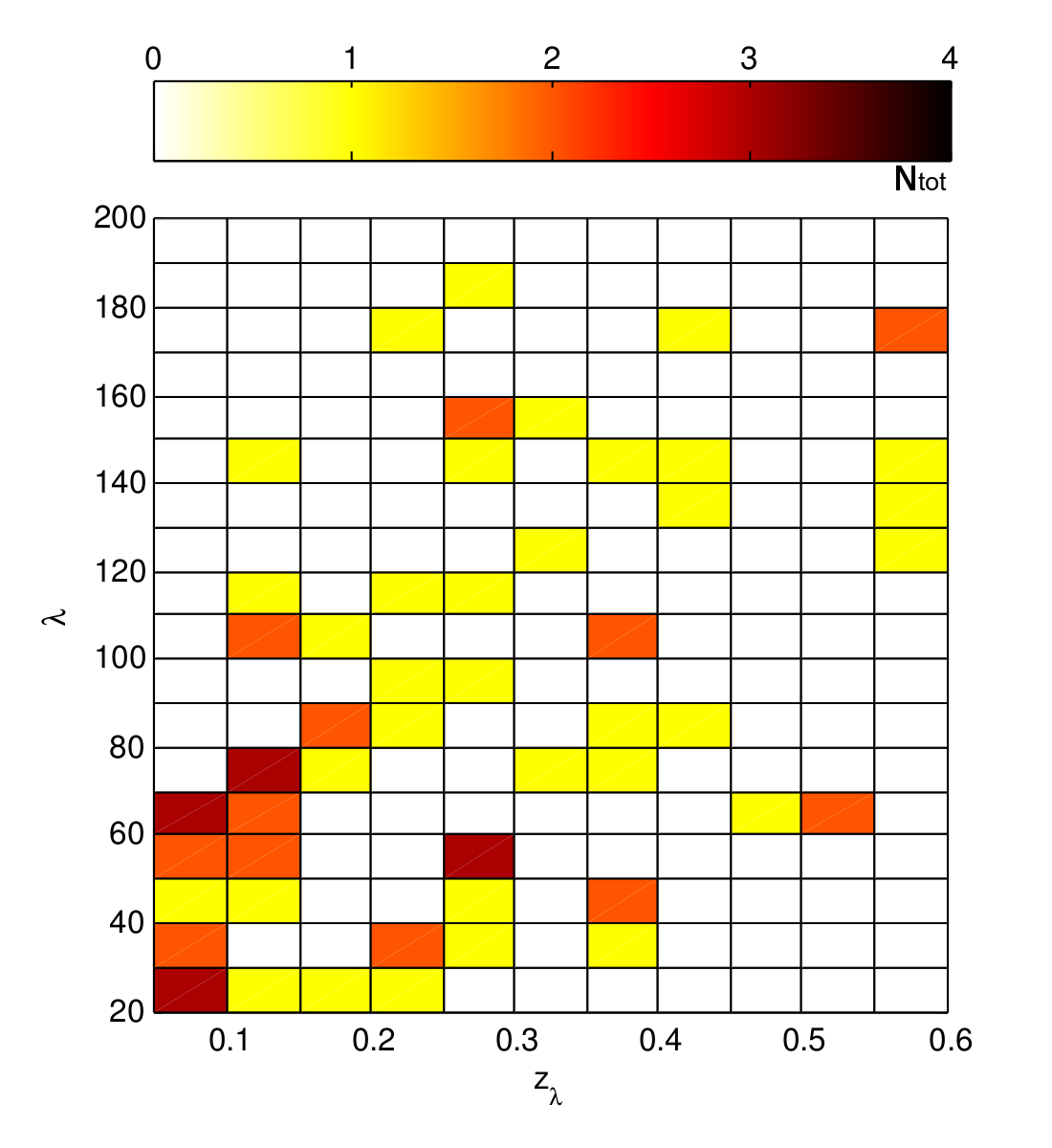} 
\caption{\textit{Redshift-richness} distribution of the 72/144 XC1$^{+}$ clusters detected as redMaPPer and corresponding to targeted  XMM observations. Same colour coding as in Fig. \ref{XC1asredMaPPerprop}}
\label{XC1pointed}
\end{center}
\end{figure}

\begin{figure}[t]
\begin{center}
\includegraphics[width=0.35\textwidth]{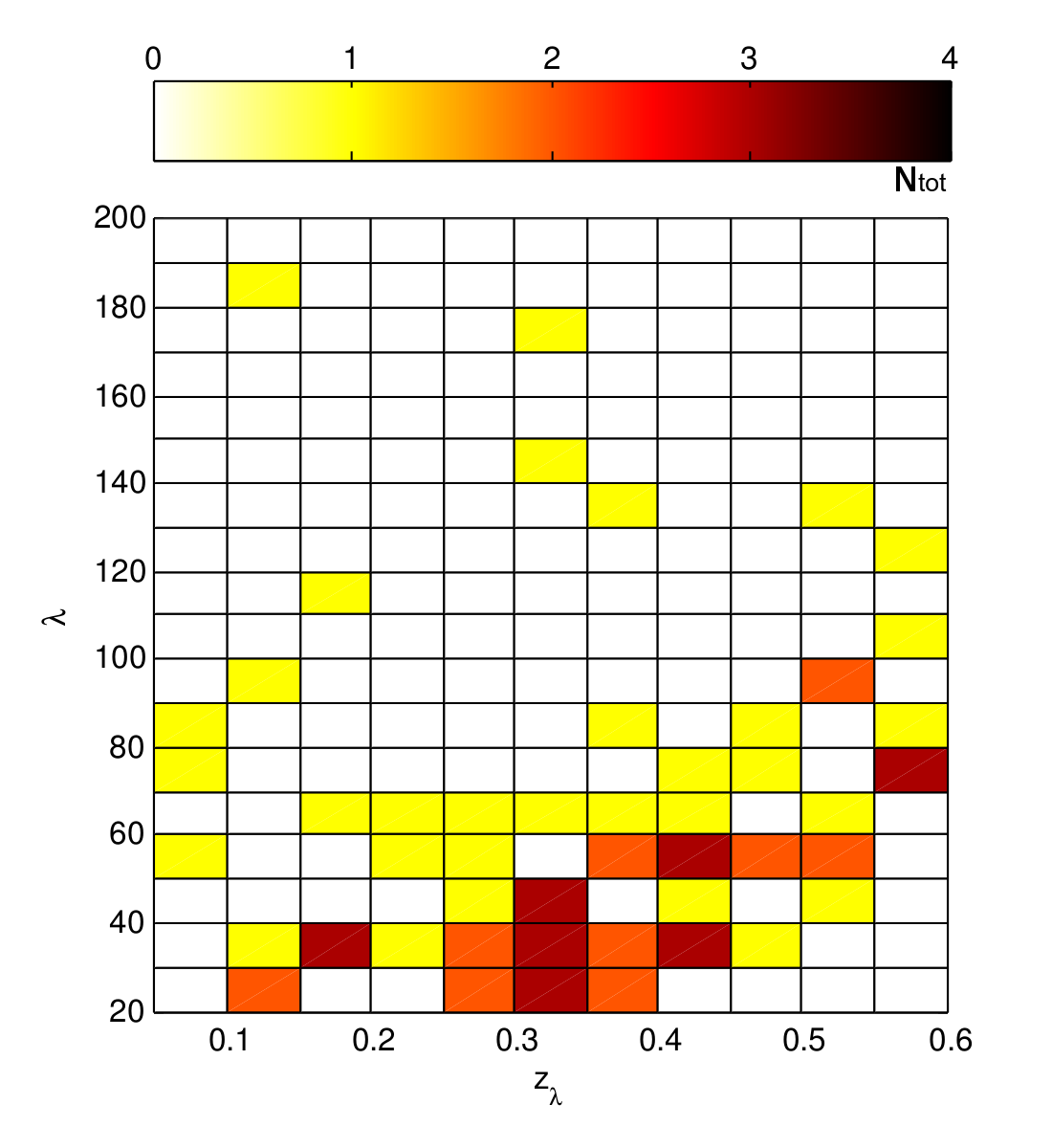} 
\caption{\textit{Redshift-richness} distribution of the 72/144 XC1$^{+}$ clusters detected as redMaPPer and corresponding to serendipitous detections in the XMM observations. Same colour coding as in Fig. \ref{XC1asredMaPPerprop}}
\label{XC1notpointed}
\end{center}
\end{figure}

Out of the 355 XC1$^{+}$ clusters falling on the redMaPPer detection area  (Fig. \ref{sky2}), 144 objects were found to have a redMaPPer counterpart. The correlation between XC1$^{+}$ and redMaPPer has been performed within a radius of 1 arcmin and yielded 121 matches to which we added 23 objects recovered after correcting for larger positional offsets (Sect. 4.2.4). We consider that this final set of matches constitutes the most realistic common X-CLASS--redMaPPer cluster sample. Their \textit{redshift-richness} distribution is shown Fig. \ref{XC1asredMaPPerprop}. 
Compared to the distribution of the full redMaPPer sample (Fig. \ref{redMaPPerprop}), we observe an overdensity of low-$z$ poor objects (nearby groups) and of rich clusters at any redshift; both can be attributed to the presence of pointed observations.
Indeed, a clear trend appears when splitting the matches into two sub-samples: (i) 72  XMM clusters are detected within an off-axis $\leqslant3\arcmin$ (assumed to be `pointed clusters', Fig. \ref{XC1pointed}) and (ii) 72 XMM clusters found beyond an off-axis $>3'$ (classified as serendipitous detections, Fig. \ref{XC1notpointed}). \\
Furthermore, we compared our eye-ball distance estimates (Sect. 2.1) for the matched XC1 with the redMaPPer photometric redshifts. We found that more than 90\% of the NEARBY objects have  $z_{\lambda}<0.4$ and more than 75\% of the XC1 classified as  DISTANT have  $z_{\lambda}>0.4$. 

 A summary of the detection statistics is presented in Table \ref{finalstat}, and
we examine below the X-CLASS clusters found not to have have any optical counterpart

\subsection{What are the X-CLASS clusters not having a redMaPPer counterpart?}

For 211 XC1$^{+}$ clusters, no redMaPPer counterpart was found. From the NASA Extragalactic Database (NED), however, information is available for a significant fraction of them: either they are classified as galaxy clusters (GClstr), or  redshifts  exist for a number of individual galaxies within the cluster fields. We have examined each of the non-matches by splitting the sample according to the X-CLASS distance estimates.  Following the criteria defined for the XMM-LSS cluster confirmation, we declare that a cluster is confirmed either when at least three galaxies with concordant redshifts are available within a radius of 500 kpc around the X-ray centroid or when the cD has a spectroscopic redshift \citep{adami11}. We note that $z_{min-redMaPPer}$ is the lowest possible redshift value of the redMaPPer clusters: $z_{min-redMaPPer} = 0.054$. Percentages are expressed as a function of the XC1$^{+}$ cardinal (355 objects in total).

\begin{figure}[t]
\begin{center}
\includegraphics[width=0.35\textwidth]{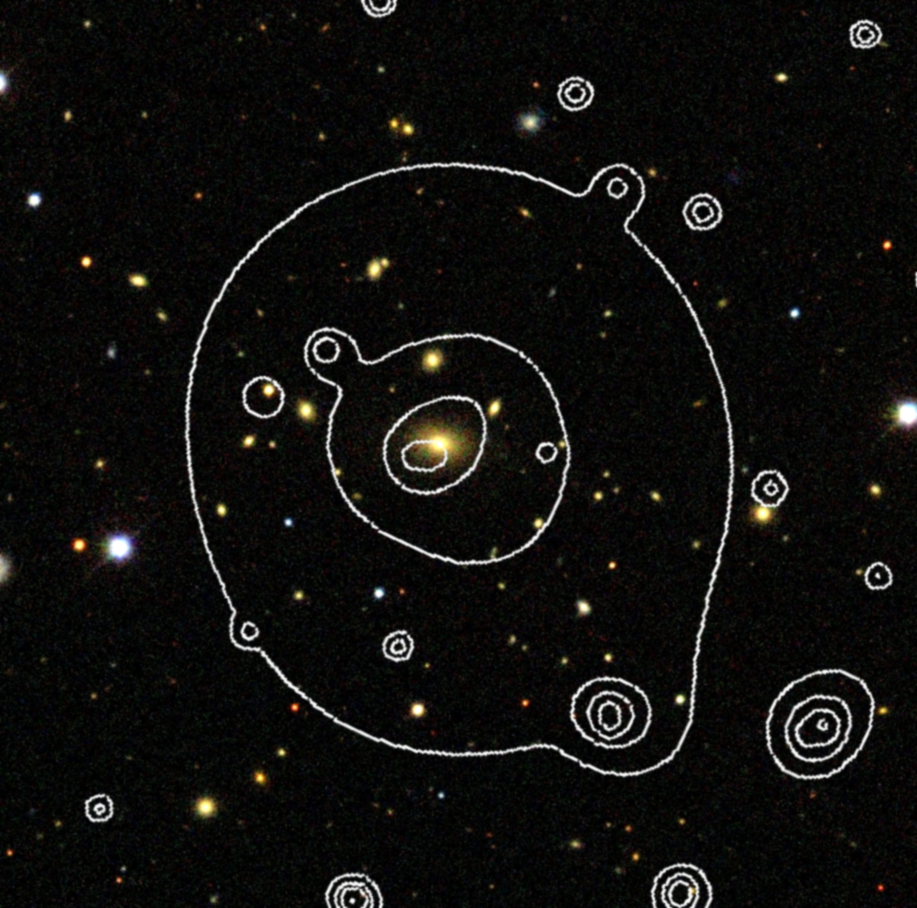} 
\caption{SDSS image and XMM contours of the NEARBY XCLASS1069 cluster not present in the $\lambda>20$ redMaPPer catalogue (field size is $6' \times 6'$). The cluster is also known as 400d J1013+4933 and has a spectroscopic redshift of  0.133. This object is actually detected by redMaPPer with a richness of 17.5 and may be a fossil group}
\label{XC1notredMaPPer}
\end{center}
\end{figure}

\begin{description}
\item[1.    ] \textbf{$\sim 28\%$} are NEARBY clusters (99 objects) and 86 of them are subsequently confirmed by means of NED information, with the following breakdown:\\
- 51 C1 have at least three spectroscopic redshifts or are found to be confirmed with spectroscopy in the literature (GClstr); an example is shown in Fig. \ref{XC1notredMaPPer}; 22 of them have $z<z_{min-redMaPPer} $\\
- For 18 C1, a spectroscopic redshift is available for the cD galaxy; two of them have $z<z_{min-redMaPPer} $\\
- 19 C1  have at least three photometric redshifts or are found to be confirmed in the literature from photometric redshift information (GClstr); none of them has $z<z_{min-redMaPPer} $\\
- For 11 C1, no information was available in NED
\item[2.    ] \textbf{$\sim 31\%$} are DISTANT clusters (112 objects), and 63 of them are subsequently confirmed by means of NED information, with the following breakdown:\\ 
- 17  C1 have at least three spectroscopic redshifts or are found to be confirmed with spectroscopy in the literature (GClstr);  the median redshift is 0.54, seven of them have $z<0.5$ \\
- For 19 C1, a spectroscopic redshift is available for the cD galaxy; the median redshift id 0.39, fourteen of them have $z<0.5$ \\ 
- 27 C1  have at least three photometric redshifts or are found to be confirmed in the literature from photometric redshift information (GClstr); they span the $0.25<z<1.14$ range (median z = 0.57), 11 of them have $z<0.5$ \\   
- For 49 C1, no information was available in NED
\end{description}

Each DISTANT C1 not identified with a redMaPPer and not already spectroscopically confirmed from NED was subsequently examined in the WISE W1/W2-bands (WISE: Wide Infrared Survey Explorer, \cite{wright10}). The purpose of this exercise was to attempt to grasp information (even limited) beyond the standard SDSS depth.
We found that one third of the clusters having a spectroscopic redshift only for the putative cD galaxy, show a conspicuous galaxy over-density. This ratio amounts to about 70\% for the C1 having at least three coincident photometric redshifts or and 60\% when there is no available information in NED. Furthermore, for more than  15\%  of the remaining objects, the X-ray emission appears clearly centred on an isolated IR galaxy that could be the cluster cD, the other cluster galaxies being too faint to be detected by WISE.  \\

\section{Discussion}

We have undertaken a detailed correlation analysis between an X-ray and an optical cluster catalogue and concentrated on the objects left out  by the procedure. The primary goal was to understand the relative impact of (i) catalogue biases (incompleteness, flaws in the cluster detection procedures, detection limits, etc.), (ii) the correlation techniques used, and (iii) intrinsic differences between the cluster populations probed by the X-ray and optical wave bands.  The ultimate aim of the procedure is to better understand how cluster samples can be reliably used in cosmological studies. 
For these  purposes, we used two catalogues that cover a large portion of the sky. After having made extensive tests, we can certify a  high degree of purity and completeness for the C1 X-CLASS and redMaPPer catalogues. In practice, the exercise turned out to require a huge amount of interactive work, i.e. visual inspection of X-ray--optical overlays for all non-primary matches, in order to establish an empirical classification of the situations encountered.

\subsection{Summary of results}
After having carefully defined the regions of overlap between the two catalogues, 
the first step was to choose an adequate correlation length. Although a physical radius would sound quite legitimate, we used a fixed angular scale of one arcmin, mainly considering the mean size of the clusters involved in the study, the limited XMM FOV, and the high density of (extended + point-like) X-ray sources, along with the good XMM positional accuracy. By  screening the X-ray--optical overlays, real associations that were found to exist beyond this radius were a posteriori added to the matched sample, and the reasons for the observed large offsets were in turned registered. In a second step, the screening analysis enabled us to define two types of categories for the unmatched sources: those due to detection pipeline shortcomings and those intrinsic to the X-ray and optical cluster populations as probed by the XMM 10ks archival data and the SDSS depths. In the regions of overlap, the X-CLASS and redMaPPer sub-samples involve 355 and 270 clusters, respectively. We can summarise our results as follows:

\begin{itemize}
 \item[•] {\em Global matching rates}. After the first correlation step, some 34\% of the redMaPPer and 26\%  of the X-CLASS XC1$^{+}$ clusters\footnote{28\% for the initial XC1 sample} were found to have an X-ray or optical cluster-counterpart, respectively. Taking correlations a posteriori recovered into account (see description below), the final matching rates  amount to 53\% and to 40\%.
\item[•] {\em Pipeline features}. The main causes of technical mismatch can be ascribed to (i) for the X-ray pipeline, its inability to characterise very bright nearby clusters filling most of the XMM detector or strongly peaked (11\% of the X-ray clusters) and (ii) for the optical pipeline, the misidentification of the cluster cD galaxy (and thus miscentring of the cluster) either because the object is not in the input catalogue (saturated or masked source) or because its colour is not compatible with a red sequence galaxy  (8\% of the optical clusters). To this must be added a few cases, generally mergers showing a complex morphology, where the cluster centre is not easily identifed both in the X-ray and optical. Those cases were then attached to the "matched" sample of clusters, that is 144 objects in total. To this, must be added clusters with a centrally peaked emission, mostly well known cool-core clusters, possibly hosting a central AGN (2\%).  
\item[•] {\em We further  correlated the optical clusters with the full X-ray catalogue} containing weak cluster candidates (C2 objects), point sources, and sources that are too weak to be characterised (either as point or as extended sources). While very few redMaPPer clusters  (5\%) were found to be a C2 source, 7\%  of them are associated - within a radius of $1\arcmin$ - with an X-ray source that can be unambiguously flagged as point source; furthermore, only five of these sources are found within $10\arcsec$ of the optical centre. From this, we can infer that at our working sensitivity level ($\sim 5~ 10^{-15}$ \flux\  for point sources) and within the redshift range probed by the redMaPPer, the masking of the ICM emission by a central AGN is a rather infrequent situation.  Moreover, 10\% of the redMaPPer are associated with a weak source ({\tt detection\_likelihood > 15}), which is too faint to be characterised. This rate increase to 19\% when extending the correlation to {\tt detection\_likelihood < 15} sources, but many of these sources are spurious.
\item[•] {\em redMaPPer clusters not coinciding with any detected X-ray emission}   represent 13\% of the redMaPPer population.
\item[•] {\em X-CLASS clusters with no redMaPPer counterpart} can be split into two subsamples: (i) those brighter than the DSS plate limit ($z<0.3-0.4$, 97 objects), where 87\%  are very likely to be real clusters according to the information available in NED (for the remaining 13\%, no information was available); (ii) those fainter than the DSS plate limit ($z>0.3-0.4$, 112 objects), where some 15\%  are confirmed clusters according to the information available in NED. For the remaining ones, only partial information or no information was available in NED: inspection of the corresponding  WISE images shows a significant galaxy over-density for more than 50\% of them.
\end{itemize}

\noindent
The main outcomes of the study are illustrated in Fig. \ref{summary-match} and statistics summarized in Table \ref{finalstat}. The redMaPPer richness appears to be a well-behaved  decreasing function of the X-ray detection likelihood: very rich clusters, detected but not analysed, either C1 or C2 populations, point sources, weak uncharacterised emission, or no X-ray emission at all.  The clusters not analysed by Xamin are mostly nearby rich objects, and redMaPPer clusters associated with weak uncharacterised X-ray sources lie mainly close to the optical detection limit.
We also discuss below some of our findings.

\subsection{Discussion}

The fraction of redMaPPer having an X-CLASS counterpart is comparable to that of the X-CLASS clusters associated with a redMaPPer (50-40\%) once pipeline shortcomings are accounted for. Nevertheless, one should keep in mind that even though both catalogues also show a similar cluster density (a few objects / \dd), the two samples probe very different redshift ranges: while the X-CLASS safely detects clusters out to $z\sim 1$, redMaPPer is limited to $z \leq 0.5$.\\
The fraction of undetected clusters that the correlation analysis enabled us to ascribe to shortcomings of the X-ray or optical pipelines is rather low, typically around 10\% of the total detections, and correspond to well-identified features of both pipelines. At least for the X-ray side, the missed bright clusters are straightforwardly recovered by eye. That huge X-ray clusters (possibly hosting a strong central source) do not pass the C1 selection comes from the statistical model used (Poissonian regime, adapted for low-surface brightness objects) and because of the area left for the background estimate that is too small.  After correcting for these obvious pipeline shortcomings, our study shows that no bright or rich, hence massive, clusters are missed by either wave band beyond the very local universe ($z> 0.05$). This is conspicuous in Fig. \ref{summary-match}, where nearly all redMaPPer clusters having no X-ray (yellow stars) are located within a thin stripe following the redMaPPer detection limit (F[richness, redshift]). Likewise, that we deliberately do not exclude targeted XMM clusters in the X-CLASS catalogue artificially raises the fraction of nearby or apparently bright (e.g. cool-core) objects.\\
Compared to the nominal X-CLASS C1 population, the fraction of C2 clusters identified with a redMaPPer is much lower. By definition, the C2 population is fainter than the C1 (clearly visible in Fig. \ref{summary-match}) and allows for some 50\% contamination by misclassified point sources.  Consequently, the low matching rate between C2 and redMaPPer both reflects that about half of the C2 are not clusters and suggests that, because they are low-luminosity objects, they probably host less prominent red sequences of galaxies. Furthermore, that so few redMaPPer are associated with a truly X-ray point source is informative as to the occurrence of a central X-ray bright AGN in nearby clusters of moderate richness. \\ 
The stacking analysis of the redMaPPer not associated with any X-ray source and falling within an off-axis of 9 arcmin, shows a $\sim$ 1-2$\sigma$ detection out to 0.5 Mpc (of the order of 0.003 c/s integrated out to 0.5 Mpc in [0.5-2] keV). Some emission is conspicuous between $0.5-1$ Mpc, which may be partly attributed to inaccurate optical positions or to fore- and background X-ray sources. After removing four redMaPPers out of the 15, the signal becomes insignificant. We therefore cannot exclude that some of the remaining redMaPPer objects are simply filaments seen in projection along the line of sight and interpreted as clusters in the photometric-redshift space. For comparison, the stacking of the seven clusters associated with sources having 0<{\tt detection\_likelihood}<15 shows a rather convincing emission profile out to 1 Mpc. Regarding the  redMaPPer falling at off-axis angles between $9\arcmin-12\arcmin$, a  1-2 $\sigma$ emission is again observed, but the situation is less clear (as conspicuous from the control random sample). This could be explained by faint undetected extended or point-source emission washed out because of the strong blurring of the XMM PSF beyond $10\arcmin$.\\  

By construction, the C1 population has a very low degree of contamination by non-cluster sources and subsequently underwent a dedicated screening. For this reason, the C1 NEARBY clusters having no redMaPPer counterpart (30\% of the C1 population, for which 90\% seem to be confirmed using literature data) deserve special attention. We subsequently correlated the unmatched XC1 clusters with the deeper redMaPPer (unpublished) catalogue extending down to a richness of 5, using a  search radius of 1 arcmin. In this way some 50\% of the unmatched XC1 were found to have an optical counterpart (53/99 and 42/112 for the NEARBY and DISTANT classes, respectively): an example is shown in Fig. \ref{XC1notredMaPPer}. This brings the percentage of XC1  having a redMaPPer counterpart to 64\%.  
The NED information and inspection of the WISE images independently confirmed more than 60\% of the DISTANT C1 not found by redMaPPer ($\lambda >20$). At this stage, it is difficult to be more conclusive as to the nature of these DISTANT objects, since we reach the limit of the comparison study in terms of survey depths. (WISE is supposed to detect only massive clusters out to $z\sim 1$ (\cite{gettings12}).)\\
We defer the in-depth study  of the  XC1 subsample having no  $\lambda >20$ redMaPPer counterpart  to a future paper (II) in which we shall explore the joint X-ray/optical multi-parameter space, allowing, say, for the galaxy colour criteria  to be relaxed.\\

\begin{figure}[t]
\begin{center}
\includegraphics[width=0.5\textwidth]{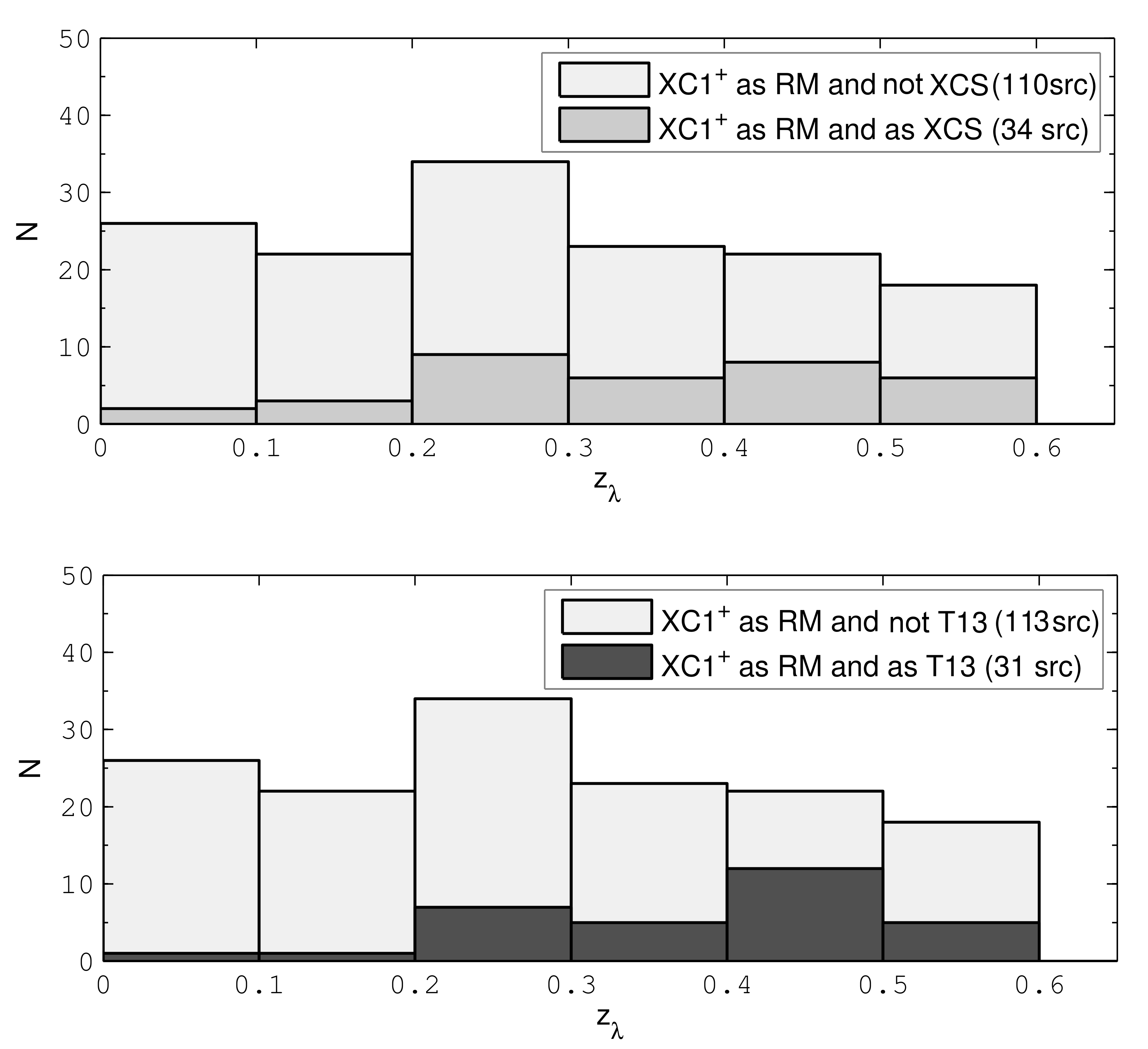} 
\caption{Comparison between the XC1$^{+}$ and XCS (top figure) and T13 (bottom) catalogues over the redMaPPer area, as a function of photometric redshift $z_{\lambda}$. The dark histograms show the redMaPPer clusters identified as XC1$^{+}$ AND XCS (or T13). The light grey histograms stand for the additional redMaPPer clusters identified with XC1$^{+}$ alone.  (144 clusters in both figures).
}
\label{XCST13}
\end{center}
\end{figure}

\begin{figure*}
\begin{center}
\includegraphics[width=1\textwidth]{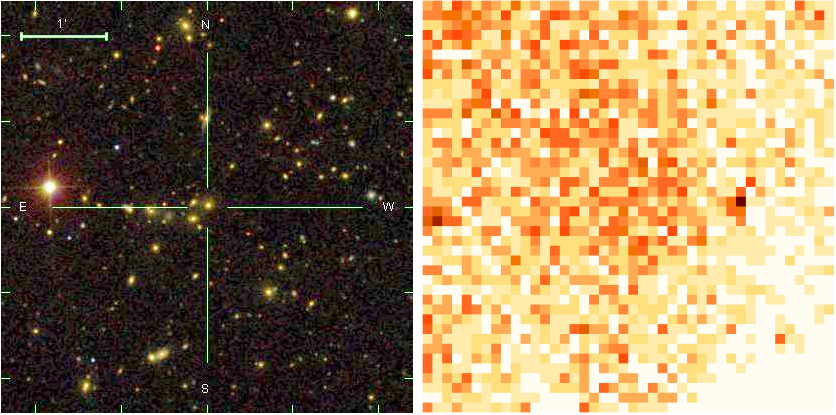} 
\caption{Optical and X-ray images of the cluster conspicuous in Fig. \ref{summary-match} at redshift = 0.24 and richness = 77, as not detected by Xamin. Marginal X-ray emission is present for this object located close to the edge of the XMM FOV, hence not entirely covered by the three detectors  (sharp discontinuities are visible on the image); this cluster was later removed from the stacking analysis of Fig. \ref{noXrand}. Centring and scale are the same for both images. }
\label{bordercase}
\end{center}
\end{figure*}

\begin{sidewaysfigure*}
   \centering
\includegraphics[width=1\textwidth]{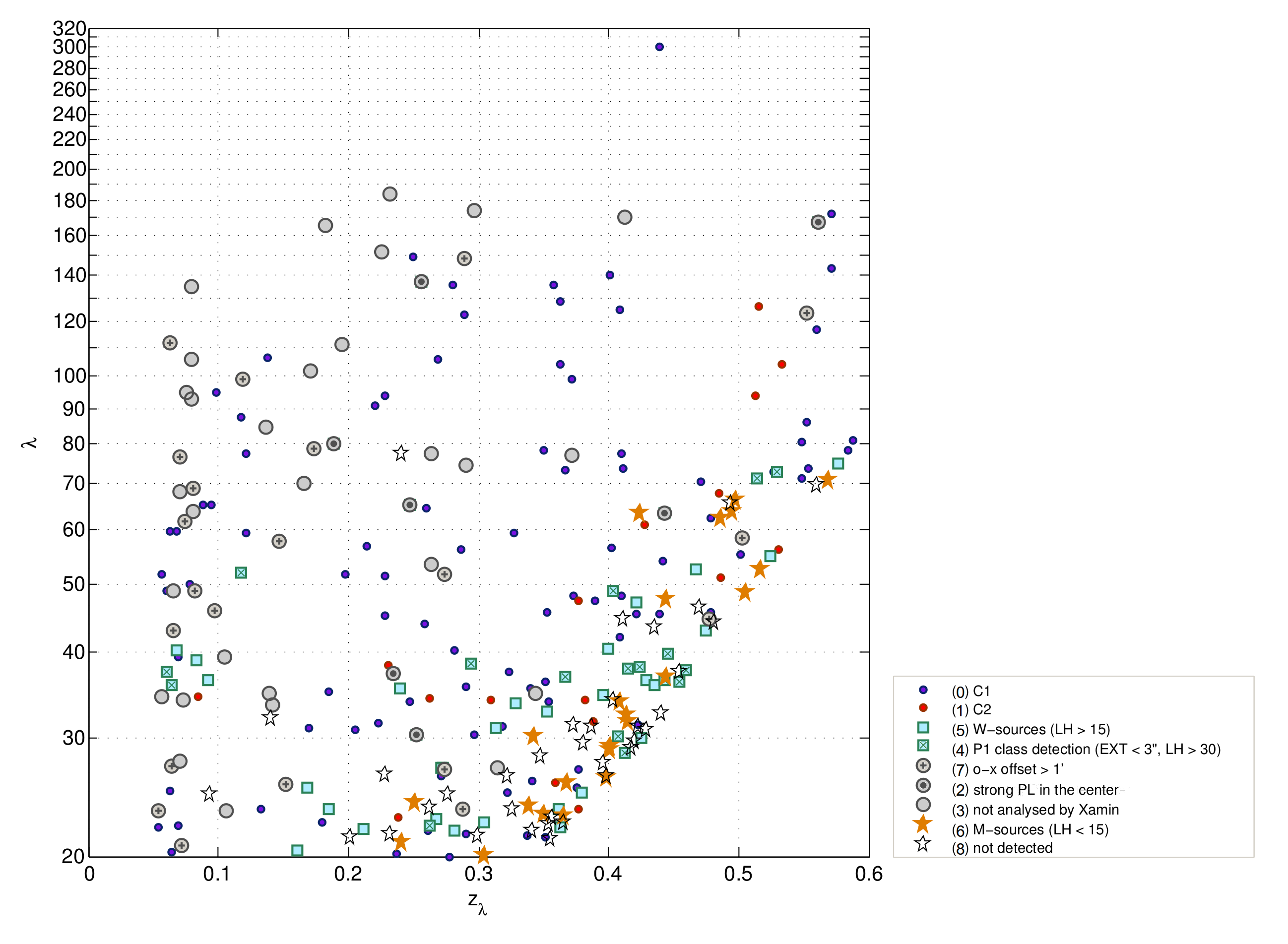} 
\caption{\textit{redshift-richness} diagramme summarizing the X-ray properties of the redMaPPer clusters (sample OPT$\rightarrow$X). Matching categories are numbered following the list presented in Sect. 4.2. The `undetected' cluster located at redshift = 0.24 and richness = 77 is shown in Fig. \ref{bordercase}}.
\label{summary-match}   
\end{sidewaysfigure*}

The \textit{redshift-richness} distribution of the  matched clusters reveals significant differences when    splitting the sample between targeted and serendipitous clusters (Figs. \ref{XC1notpointed} and \ref{XC1pointed}, respectively). Given the very large number of clusters ($\sim 200$ ) that have now been pointed by XMM (and being most of the time as conspicuous in the figure, the brightest objects in any redshift slice), one cannot simply exclude these objects from the final cluster samples,
as usually done. This is especially critical if these samples are used for subsequent scaling-relation or cosmological analyses. As a test, we correlated the X-CLASS clusters matched with redMaPPer with two other cluster catalogues also extracted from the XMM archive (XCS and T13),  still using a correlation length of one arcmin. Neither XCS nor T13 includes targeted clusters, and we have not investigated the overlapany further between the initial sub-sets of selected XMM observations from which the three catalogues were drawn. The results are shown in Fig. \ref{XCST13}, where we find that only 22\% and 24\% of the XCS and T13  clusters are in common with the X-CLASS-redMaPPer sample. Removing the targeted clusters from X-CLASS raises these percentages to 37\% and 41\%. It is not the purpose of the present paper to perform a comparison between the various serendipitous XMM cluster catalogues - all the more so since XCS and T13 did not make their selection function explicit - but from the present plots, it is clear that the differences are significant. 

\section{Conclusions}

We have undertaken a non-blind generic comparison between two cluster samples defined in the X-ray and optical wavebands, concentrating on the left-out. The overlap samples involve  some 270 (optical) and 355 (X-ray) objects and have well-defined selection functions, which does not a priori imply a one-to-one correspondence: the  C1 clusters constitute a  high X-ray surface brightness sample out to a redshift of $z<1.5$, the redMaPPer objects are red-sequence clusters limited to $z \sim 0.5-0.6$. 
The analysis of the non-matched objects has benefited from extensive human inspection. Main conclusion is that we found no evidence for any optically rich cluster to be devoid of X-ray emitting gas and vice versa.
  For SDSS imaging, and given the observational depth of the XCLASS catalogue, we find that all $\lambda > 80$ galaxy clusters in the redshift range z<0.6 are detected by both algorithms.  This corresponds roughly to M200c $\sim 4\times 10^{14} h^{-1} M_{\odot}$. This is a reasonable match to the X-ray luminosity redMaPPer detection threshold of $\sim 2\times 10^{44}$ ergs/s derived in \citet{rozo14b}. Mass detection limits will be discussed in Paper III, which will present the X-ray/optical scaling relations.\\

The comparison has not only usefully enlightened a few shortcomings of both detection methods but also, most importantly, enabled us to pinpoint key issues for future cluster science.
It is difficult  to define a unique matching radius that takes all specificities of the two samples into account, both from the instrumental and from the cluster-physics points of view, hence the need for an interactive approach. Moreover, the limited XMM field of view, vignetting, and PSF clearly set practical limits to the stacking analysis. Similar to optical cluster catalogues, X-ray serendipitous catalogues show significant differences between each other. In any case, the selection functions have to be explicitly involved in the process. All these aspects have a critical impact on any X-ray--optical scaling-relation work. This leads us to stress again that cluster evolution, selection effects, and cosmology cannot be worked out independently. In Paper III, we shall present the joint X-CLASS and redMaPPer catalogue along with scaling relations.\\
This very instructive approach has only provided an overview of the difficulties and promises of dedicated   X-ray/optical cluster studies involving hundreds of objects and could be easily extended to X-ray/X-ray, optical/optical optical/S-Z (e.g. \citet{rozoetal14}), etc... comparisons. The current lack of redshifts for a large number of the southern X-CLASS clusters is being addressed by systematic  multi-band observations with the GROND instrument on the MPG/2.2m telescope at La Silla (\cite{greiner08}), to obtain images and reliable photo-z for a large portion of the catalogue in the southern sky (Clerc et al. in prep).
The next steps are obviously to extend the comparison to optical catalogues based on other detection methods and going deeper in the optical and IR wavebands, as well as using ancillary, deeper XMM observations when available. It is nevertheless anticipated that projection effects and cluster evolution issues will get more severe with increasing redshift, hence the need for a truly multi-wavelength approach. It is also obvious that numerical simulations are to play a growing role in cluster detection and subsequent matching studies. 
 In this respect, the XXL project  provides a unique data set (Pierre et al in prep).

\begin{acknowledgements}
We are grateful to Florian Pacaud for useful discussions and to Ali Takey for providing information about his catalogue. Tatyana Sadibekova acknowledges a post-doctoral position from the Centre National d'Etudes Spatiales (CNES).This work was supported in part by the U.S. Department of Energy contract to SLAC no. DE-AC02- 76SF00515.
\end{acknowledgements}

\bibliographystyle{aa}
\bibliography{mylib}{}

\begin{table*}
\caption[]{Summary of the correlation statistics}
\label{tab2}
\centering
\begin{tabular}{lcc}

\hline
\noalign{\smallskip}
& redMaPPer & X-CLASS\\
\noalign{\smallskip}
\hline
\noalign{\smallskip}

density & $\sim2.5/deg^{2}$ & $\sim5/deg^{2}$\\
\noalign{\smallskip}
redshift range & &NEARBY~~~~~~~~~~~~~~~~~DISTANT\\
& $0.054 < z < \sim0.6$& $0 < z < 0.3-0.4$~~~~~~$0.3-0.4 < z < 1.5$ \\
\noalign{\smallskip}
\noalign{\smallskip}
median mass range (M$_{500c}$) in unit of $10^{14}h^{-1}M_{\odot}$  & $\sim 1$ at $ z < 0.35$ and $\sim 3.5 $ at $z\sim 0.55$  &$0.2$~~~~~~~$5 $\\
& & from \citet{pacaud07}\\
\noalign{\smallskip}
overlap sample& 270 clusters & 355 clusters \\
\noalign{\smallskip}
\hline
\noalign{\smallskip}
\noalign{\smallskip}
 fraction of matched objects & $34\%$ & $26\%$\\
\noalign{\smallskip}
fraction of recovered matches & $8+11\%$ & $6+8\%$\\
(large offsets + pipeline failures) & & \\
\noalign{\smallskip}
 fraction with no counterpart at all& $13\%^{(*)}$ & $60\%^{(**)}$\\
&$^{(*)}$not matched with any type  & $^{(**)}$either not seen by redMaPPer\\
&of X-ray source&or beyond $z > 0.6$\\
\noalign{\smallskip}
\hline
\label{finalstat}
\end{tabular}
\end{table*}

\end{document}